# Accretion of Saturn's mid-sized moons during the viscous spreading of young massive rings: solving the paradox of silicate-poor rings versus silicate-rich moons.

Sébastien CHARNOZ *,1
Aurélien CRIDA 2
Julie C. CASTILLO-ROGEZ 3
Valery LAINEY 4
Luke DONES 5
Özgür KARATEKIN 6
Gabriel TOBIE 7
Stephane MATHIS 1
Christophe LE PONCIN-LAFITTE 8
Julien SALMON 5,1

(1) Laboratoire AIM, UMR 7158, Université Paris Diderot /CEA IRFU /CNRS, Centre de l'Orme les Merisiers, 91191, Gif sur Yvette Cedex France
(2) Université de Nice Sophia-antipolis / C.N.R.S. / Observatoire de la Côte d'Azur Laboratoire Cassiopée UMR6202, BP4229, 06304 NICE cedex 4, France
(3) Jet Propulsion Laboratory, California Institute of Technology, M/S 79-24, 4800 Oak Drive Pasadena, CA 91109 USA
(4) IMCCE, Observatoire de Paris, UMR 8028 CNRS / UPMC, 77 Av. Denfert-Rochereau, 75014, Paris, France
(5) Department of Space Studies, Southwest Research Institute, Boulder, Colorado 80302, USA
(6) Royal Observatory of Belgium, Avenue Circulaire 3, 1180 Uccle, Bruxelles, Belgium
(7) Université de Nantes, UFR des Sciences et des Techniques, Laboratoire de Planétologie et Géodynamique, 2 rue de la Houssinière, B.P. 92208, 44322 Nantes Cedex 3, France
(8) SyRTE, Observatoire de Paris, UMR 8630 du CNRS, 77 Av. Denfert-Rochereau, 75014, Paris, France

(*) To whom correspondence should be addressed (charnoz@cea.fr)

## ABSTRACT

The origin of Saturn's inner mid-sized moons (Mimas, Enceladus, Tethys, Dione and Rhea) and Saturn's rings is debated. Charnoz et al. (2010) introduced the idea that the smallest inner moons could form from the spreading of the rings' edge while Salmon et al. (2010) showed that the rings could have been initially massive, and so was the ring's progenitor itself. One may wonder if the mid-sized moons may have formed also from the debris of a massive ring progenitor, as also suggested in Canup (2010). However, the process driving mid-sized moons accretion from the icy debris disks has not been investigated in details. In particular, this process does not seem able to explain the varying silicate contents of the mid-sized moons (from 6% to 57% in mass). Here, we explore the formation of large objects from a massive ice-rich ring (a few times Rhea's mass) and describe the fundamental properties and implications of this new process. Using a hybrid computer model, we show that accretion within massive icy rings can form all mid-sized moons from Mimas to Rhea. However in order to explain their current locations, intense dissipation within Saturn (with $Q_p<2000$) would be required. Our results are consistent with a satellite origin tied to the rings formation at least 2.5 Gy ago, both compatible with either a formation concurrent to Saturn or during the Late Heavy Bombardment. Tidal heating related to high-eccentricity post-accretional episodes may induce early geological activity. If some massive irregular chunks of silicates were initially present within the rings, they would be present today inside the satellites' cores which would have accreted icy shells while being tidally expelled from the rings (via a heterogeneous accretion process). These moons may be either mostly icy, or, if they contain a significant amount of rock, already differentiated from the ice without the need for radiogenic heating. The resulting inner mid-sized moons may be significantly younger than the Solar System and a ~1Gyr delay is possible between Mimas and Rhea. The rings resulting from this process would evolve to a state compatible with current mass estimates of Saturn's rings, and nearly devoid of silicates, apart from isolated silicate chunks coated with ice, interpreted as today Saturn's rings' propellers and ring-moons (like Pan or Daphnis).

# 1. Introduction

Between the outer edge of the Saturn's rings and Titan (about 2.3 and 20.3 times Saturn's equatorial radius respectively) orbits the population of ice-rich inner moons with an organized orbital architecture in which the mass is an increasing function of distance (Figure 1). They are organized into two distinct families, the "small moons" interior to Mimas' orbit, with irregular shapes, and the "mid-sized moons," extending from Mimas to Rhea, with ellipsoidal shapes and average-radii ranging from 198 km (Mimas) to 764 km (Rhea) (Thomas et al., 2007, Thomas 2010). The physical peculiarities of the latter population still challenge all formation models:

a) The averaged density of the mid-sized moon system is at least 25% less than Titan's uncompressed density and much below the density expected for a Solar composition (Johnson and Lunine 2005). This suggests that the material accreted into these objects was depleted in rocks, but the mechanism responsible for that situation remains to be found. Shape data suggest that these moons still present a central concentration of material (Thomas et al., 2007, Thomas, 2010) characteristic of a partially differentiated structure, or, alternatively, of an internal gradient in porosity (e.g., Eluszkiewicz and Leliwa-Kopystynski, 1989). The geophysical evolution and current state of these objects are not well understood (Schubert et al., 2010). Contrarily to large icy satellites, accretional heating was not a significant heat source that could drive early melting and separation of the rock from the volatile phase in small satellites (Squyres et al. 1988; Matson et al. 2009). Differentiation in that class of objects is a function of their time of formation, which determines the amount of accreted short-lived radioisotopes (Castillo-Rogez et al. 2007; Barr and Canup 2008; Matson et al. 2009) and their tidal dissipation history, which is poorly constrained. Still, while the masses of Saturn's mid-sized satellites span over two orders of magnitude (from Mimas to Rhea), they all share a high-albedo (e.g., Owen et al., 1995), icy surface, implying that silicates have been efficiently and systematically removed from their surfaces and buried in their interior, by some unknown mechanism. In this respect, Saturn's icy moons are unique in the Solar System.

b) The cratering frequency at the surface of Saturn's satellites is puzzling in several respects. It has been recognized since the Voyager epoch (Smith et al., 1982; Lissauer et al., 1988; Zahnle et al., 2003; Dones et al., 2009) that the relative cratering records of Rhea and Iapetus are inconsistent if they were struck by a common population of heliocentric impactors (like comets). Indeed, since Rhea is closer to Saturn than Iapetus, the planet's gravitational focusing should produce a much larger crater density on Rhea's surface than on Iapetus' surface. However, the two moons have similar crater densities. For smaller craters, that might be a result of crater saturation equilibrium (Hartmann 1984) having been reached on the saturnian satellites, the many basins on Iapetus cannot be explained in this way. Consequently, Rhea's surface seems younger than Iapetus' surface. In addition, if the population of impactors was heliocentric and the moons had already achieved synchronous rotation, (as they seem today), a large apex/antapex asymmetry should be visible on their surfaces (Zahnle et al., 2001, 2003; Dones et al., 2009). Such an enhancement is not visible, although Schenk and Murphy (2011) have found smaller excesses of rayed craters on the leading hemispheres of several of the mid-sized moons. This is one of the reasons why it is thought that Saturn's icy satellites were hit by a population of planetocentric impactors (Zahnle et al., 2003; Dones et al., 2009) in addition to, or even instead of, a heliocentric one. Extensive crater analysis has recently brought further observational evidence that Saturn's inner medium-sized bodies were affected by at least two populations of impactors (Kirchoff and Schenk 2010). These authors also noted a contrast in crater densities between the plains and the large craters of several Saturnian satellites, especially at Tethys and Mimas. In the latter case, Schenk (2011) inferred that the crater Herschel is especially young, since it is pockmarked by only a few small craters. On the other hand, Kirchoff and Schenk (2011) noted an anomalous dearth of large craters at Mimas, Dione, and Tethys, in comparison to the frequency distribution expected in the case of a population of impactors that would have hit the satellites during the Late Heavy Bombardment [LHB] (taking Iapetus as a reference), which cannot simply be explained by viscous relaxation. Despite the increasing observational evidence that the LHB cannot account for the observed crater distribution and frequency, the origin of a separate population of interlopers, probably in the form of planetocentric objects, has never been identified. Note that works modelling Iapetus' formation (Mosqueira and Estrada, 2003a, 2003b, Mosqueira et al., 2010) have suggested that Iapetus' surface may have been impacted by proto-satellites remaining at the end of its formation. These satellitesimals could constitute a potential population of planetocentric interlopers at least for Iapetus. However it is unclear if the signs of their impacts on the surfaces of Mimas to Rhea would be still visible today. So the cratering history of Saturn's satellites and the question of the population of impactors is still an open question.

c)The most puzzling properties of these satellites is perhaps their varying silicate fractions (Fig. 2): 26%, 57%, 6%, 50% and 33% for Mimas, Enceladus, Tethys, Dione and Rhea, respectively, assuming the moons have no porosity (Matson et al., 2009). Oddly enough, the uranian satellites present a somewhat similar trend in their silicate fractions. What kind of process could produce such diversity? This problem still challenges any formation scenario and his largely not addressed. Estrada and Mosqueira (2006) suggest that a stochastic collisional capture of interlopers penetrating deep in the planet's gas poor sub-nebula may be at the origin of the varying silicate contents of the mid-sized moons. Unfortunately they do not explore this process in detail. Other published models do not address this question (see e.g. Canup and Ward 2006, Sasaki et al., 2010).

d) The current location of Saturn's mid-sized moons obeys a simple architecture where the mass is an increasing function of distance from the planet (Fig. 1). This can be interpreted as the signature of tidal forces driving the outward migration of the satellites at a rate proportional to their mass. It depends on the planet's dissipation factor, hereafter labelled $Q_p$. In the sole presence of Saturn's tides, the satellite's semi-major axis should evolve according to the following law (Goldreich and Soter, 1966):

$$\frac{da_s}{dt} = \frac{3k_{2p}M_s G^{1/2} R_p^5}{Q_p M_p^{1/2} a_s^{11/2}}$$

**Eq. 1**

with $M_p$, $k_{2P}$ and $R_p$ standing for the planet's mass, tidal Love number, and equatorial radius, respectively, while $M_s$ and $a_s$ stand for the satellite's mass and semi-major axis, and $G$ is the universal gravitational constant. $Q_p$ is usually assumed to be not less than 18,000 for Saturn with $k_{2P} \sim 0.341$ (Gavrilov & Zharkov 1977, see also Dermott et al., 1988, Sinclair 1983, Meyer and Wisdom 2007). However with this canonical value it is not possible to easily explain the current and very ordered architecture in mass of Saturn's satellites (see Appendix A of the present paper) if one assumes they all formed at the same moment (apart from doing a very fine, unlikely adjustment of the initial positions, so that today they seem regularly ordered while they were not in the past). Of course we have only very few constraints on $Q_p$ and it might have changed with time. We remind our reader that the canonical lower bound of Saturn's Qp is inferred from the orbital evolution of Mimas to its current location based on four assumptions (and neglecting the torque exerted by the rings) (Goldreich, 1965): Mimas formed (i) just outside of Saturn's synchronous orbit (~ 113,000 km from Saturn's center); (ii) 4.5 Gyr ago; (iii) with its current mass and (iv) simultaneously with the other satellites. Under these particular assumptions, $k_{2p}/Q_p$ was found to be lower than $2\times10^{-5}$ (Goldreich 1965, Goldreich & Soter 1966). Examination of the Tethys-Mimas resonance yields $Q_p \geq 18,000$ (Sinclair, 1983) for $k_{2P}\sim0.341$ (Gavrilov & Zharkov 1977) also under the four aforementioned assumptions. For comparison, Dermott et al. (1988) finds $Q_p \geq 16,000$. These assumptions are, in fact, poorly constrained and justified. We will show below, in particular, that these assumptions are incompatible with formation of the satellite at the rings' edge (because in this case the satellite's age increases with the body's mass, see section 3). Recently Lainey et al. (2010) re-examined Saturn's $Q_p$ on the basis of astrometric observations of Saturn's satellites spanning over one century. They found $Q_p \sim 1680$ (assuming $k_{2p}\sim0.341$), implying a much stronger tidal dissipation in the planet's interior than previously thought, and in turn, a much faster orbital expansion. In the present study both this value and the "traditional" lower bound of $Q_p$ will be considered.

### 1.1 Linking moon formation to ring formation

#### 1.1.1 How to form satellites?

The formation of Saturn's satellites in Saturn's sub-nebula has been investigated by several authors exploring different structures of the Saturn's sub-nebula (see, e.g., Mosqueira and Estrada, 2003a,b; Estrada and Mosqueira 2006; Canup and Ward 2006; Barr and Canup 2008; Estrada et al., 2009; Johnson and Estrada 2009; Mosqueira et al., 2010; Sasaki et al., 2010). Whereas there is some understanding of the fact that Saturn's moons should be smaller than Jupiter's and that the saturnian system may be dominated by a single large moon (Titan) rather than a collection of similar-sized moons (the galilean satellites), no model self-consistently explains the origin of the mid-sized icy satellites and, in particular, their varying silicate contents. For example, models of moon formation in a gas-rich saturnian nebula fed by planetesimals from the solar nebula (Mosqueira et al., 2010) imply a monotonically increasing fraction of silicates closer to Saturn (because the closer to Saturn, the denser the nebula and thus denser material is more easily trapped there). Otherwise, most satellite accretion scenarios occurring in Saturn's nebula, proposed up to now, simply do not address this problem that is very specific to Saturn's mid-sized moons and is not easily tied to the Saturn's nebula properties (see, e.g., Canup and Ward, 2006, Mosqueira et al., 2010). Mosqueira and Estrada (2003a,b) suggested that stochastic collisions with planetesimals may be at the origin of these peculiarities. However, this idea has not been investigated so far.

Recently a new mechanism (Charnoz et al., 2010) was proposed to explain the origin of Saturn's smallest moons orbiting between the outer edge of Saturn's rings and below Mimas' orbit (namely, Atlas, Prometheus, Pandora, Epimetheus, and Janus): the spreading of ring material beyond Saturn's Roche limit, resulting in the formation of clumps that coalesce to produce today's moonlets. The resulting orbital architecture of the small moons, as well as their size distribution, was reproduced (see Fig. 3 and Fig. 4 of Charnoz et al., 2010). This mechanism of moon formation is in fact more general than the sole case of Saturn's small moons moons: the formation of the Earth's Moon can be described by the same process (Charnoz et al., 2010), namely accretion of a proto-satellite at the Roche limit of the proto-lunar disk, supporting the idea that satellite formation at the edge of circumplanetary debris disks within the planet's Roche limit may be a general mechanism of satellite formation. In this perspective, we need a disk to be the progenitor of a moon system. So then comes naturally the question of the origin of Saturn's rings… a long standing question !

## 1.1.2 How to form rings?

The destruction of a large satellite close to Saturn has long been proposed as a possible origin for the ring system (Harris, 1984; Charnoz et al., 2009a, 2009b; Canup 2010). Recently, Canup (2010) has shown, using SPH simulations, that a Titan-sized differentiated moon drifting toward Saturn (because of Type 1 migration in Saturn's nebula) has its icy mantle progressively "peeled off" by tides as it approaches the planet below ~3 Saturn radii. The icy mantle is ground into small pieces and remains in orbit while the silicate core is lost into the planet because of tides with the disk. This process may nicely explain the apparent absence of silicates from Saturn's main rings (Cuzzi and Estrada, 1998; Poulet et al., 2003; Nicholson et al., 2005; Cuzzi et al., 2009), however it may be still subject to questions concerning its initial conditions (the possibility for a differentiated Titan-sized satellite at this time, the nebula model that fuels inward migration, Mosqueira and Estrada 2011). However the main idea – the tidal destruction of a differentiated satellite – still remains the best model today to explain the very existence of silicate poor rings. The ring's progenitor must be sufficiently massive to suffer type I migration in Saturn's gas nebula (in order to penetrate inside Saturn's Roche zone), so its mass is intimately linked to the nebula's surface density. In Canup's work this translates into a Titan-sized satellite (see e.g. Canup and Ward 2006, Canup 2010), whereas with different assumptions on the Saturn's nebula, a Rhea-sized body may also suffer migration (see e.g. Mosqueira et al., 2010). In the present work we will simply assume that the ring's progenitor is at least a few times Rhea's mass and we do not make any particular assumption on Saturn's nebula. Note that we cannot exclude that the ring progenitor could be a massive (> 1000 km) unique heliocentric interloper. However, we tend to disregard such a scenario as it has been shown that if Saturn's rings origin is linked to cometary implantation (during the LHB and coming from an early massive Kuiper Belt) then all four giant planets may be expected to show massive ring systems (Charnoz et al., 2009) and they would have formed from the debris of a numerous population of small bodies, in which ice and silicates could be intimately mixed (see section 4.1).

Canup (2010) also suggests that the same mechanism that formed Saturn's smallest moons may have formed all of Saturn's mid-sized moons up to Tethys (e.g., Mimas, Enceladus, and Tethys, but not Dione and Rhea), extrapolating the work of Charnoz et al. (2010) initially intended explain the small moons population, to very high mass objects. In this scenario, the origin of Saturn's moons is intimately linked to the rings' origin. However, the process by which such large moons may accrete was not described. Also,Canup (2010) does not provision any explanation for Dione and Rhea, which suggests that in that framework, the inner moons would stem from two different accretion processes.

Other scenarios have suggested that a differentiated satellite close to Saturn may have been hit by a comet (coming from the primordial Kuiper belt) during the Late Heavy Bombardment , resulting in catastrophic disruption of the moon (Charnoz et al., 2009a, 2009b). This scenario has the advantage of explaining naturally why Saturn is the only planet to have massive rings (due to the relative position of its Roche limit to its synchronous orbit, Charnoz et al., 2009a). A Mimas-sized satellite would be expected to suffer about one catastrophic impact during the LHB (Charnoz et al., 2009a) so any moon larger than 100 km radius present at the time of the LHB close to Saturn's synchronous orbit might have been destroyed. Based on these arguments, Charnoz et al. (2009a, 2009b) speculate that, if the moons were differentiated, their mantles would be preferentially disrupted (because of shock-wave reflection at the core/mantle boundary) and scattered into orbit (as was the case for the proto-Earth during the impact that resulted in the formation of the Earth's Moon; see, e.g., Canup 2004) and might result in the formation of an icy ring. In the current state of our knowledge, this is still a speculative scenario and requires further study. We also note that the survival of a large moon close to Saturn's Roche limit up to the onset of the LHB event, occurring about 700-800 My after Saturn's formation would require a high value of $Q_p$ (Charnoz et al., 2009a) and is a strong constraint on this model.

Whatever the process that formed Saturn's rings, it is possible that the rings were initially very massive (Salmon et al., 2010), i.e. about 100 to 1000 times the present mass of Saturn's rings (which is thought to be about Mimas' mass or perhaps a few times larger — see, e.g., Charnoz et al., 2009b; Robbins et al., 2010). Then they would viscously evolve toward their present-day mass. Indeed, Salmon et al. (2010) have recently shown that, due to gravitational instabilities, the initial rings would initially spread very rapidly and would lose most of their mass either through their inner edge (to the planet's atmosphere) or through their outer edge (through the Roche limit). Progressively, the ring stabilizes into a state of marginal gravitational stability (with the Toomre parameter close to 2, see Salmon et al., 2010). The resulting asymptotic value of the disk's mass would be about Mimas' mass (as predicted by the analytical calculation of Salmon et al., 2010; see their equations 22 and 23), in qualitative agreement with the estimates of the current mass of Saturn's rings (Charnoz et al., 2009b).

## 1.2 Purpose and structure of the paper

The purpose of this paper is to investigate the process of moon accretion at the edge of an initially massive saturnian ring either made of pure ice (section 3) or containing some silicate chunks (section 4). We do not focus on a particular model of Saturn's rings formation (either in the Saturn's nebula or during the LHB, either through tidal destruction of a satellite or through a cometary impact). We simply study the evolution of an initially massive disk, giving birth to satellites through gravitational instability, in a *gas-free* environment in order to understand the basic physical properties of this process. We wish to answer the following questions: What is the mass of the moons that could be formed at the rings' edge? Is it possible to reproduce the current orbital architecture and size-distribution? How old are Saturn's satellites? Is it possible in this framework to reproduce the current mass of Saturn's rings?

In Section 2 our model is briefly described, and in section 3 our results concerning the moon's accretion and physical and chronological properties are discussed. Finally, in Section 4, we wish to investigate the difficult problem of the variations in silicate content across the mid-sized moon system. Canup (2010) hypothesized that the ring progenitor was fully differentiated and the silicate phase was lost in the planet. However, in case of full differentiation, only the water-dominated moons could be formed in that context (e.g., Tethys, possibly Mimas), while a different avenue needs to be considered for the rock-rich satellites. We demonstrate that if the ring progenitor was only partially differentiated (as suggested by some authors, see e.g. Mosqueira and Estrada 2011), or if, about 1% of its silicate-core mass remained in the icy debris disk after the tidal stripping, it becomes possible to devise an accretion framework that explains simultaneously the existence of (a) mid-sized moons with a variety of silicate content (b) small moons and rings devoid of silicates. Indeed, in Section 4 we introduce a scenario in which icy material accretes at the surface of large silicate chunks (> 10-100 km) inside Saturn's Roche limit. We call this process "heterogeneous accretion". We show that heterogeneous accretion leads to moons that are already chemically differentiated without the need for radiogenic heating. In this framework, the rock mass fraction may vary stochastically between moons giving a potential explanation to this long-time riddle.

## 2. The model

We use a hybrid computer model called "HYDRORINGS" described in Charnoz et al. (2010) and in Salmon et al. (2010). HYDRORINGS couples a 1-D hydrocode for the rings' surface density evolution using a realistic viscosity model (accounting for enhanced viscosity when the disk is self-gravitating, see Daisaka et al., 2001) as well as the moons' gravitational torques (using the formalism of Meyer-Vernet and Sicardy, 1987) that is considered for all first-order mean-motion resonances. The 1-D ring density evolution is computed with a second-order solver using a staggered grid. The disk material grain density is always 900 kg/m$^3$, compatible with an ice composition. As described in Charnoz et al. (2010) satellites are formed when any ring material crosses the planet's Roche limit for ice (computed as $R_r$=140,000 km). At every time step, all the ring mass inside the grid-cells (of the simulation) that are located beyond the $R_r$ are converted into satellites (one per grid-cell), while the corresponding grid-cells are emptied of their material content (see Charnoz et al., 2010 for more details). Once a satellite is formed, its mass, eccentricity and semi-major axis are tracked ($m_s$, $a_s$ and $e_s$ respectively). To compute the satellite's radius, we assume that its average density is always 1000 kg/m$^3$ for simplicity. Due to the excessively short orbital periods of Saturn's moons compared to the age of the Solar System (from ~15 hours at the ring's edge to about 400 hours for Titan), it is not possible to use a direct N-body simulation of moon's orbital evolution. As in Charnoz et al., (2010), we instead use a very simple orbital model that accounts for tidal interactions with the planet and with the rings. However, we introduce a new effect that was not treated by Charnoz et al. (2010): the satellites' internal dissipation, which may be a strong effect in moons on eccentric orbits. The semi-major axes of satellites ($a_s$) expand because of the ring's torque $\Gamma_s$ and planet's tides but that evolution may be counteracted by dissipation in the satellites. Then the satellites' semi-major axes and eccentricities, $e_s$, evolve according to (Kaula 1964, Peale and Cassen 1978):

$$\frac{da_s}{dt} = \frac{3k_{2p}M_s G^{1/2} R_p^5}{Q_p M_p^{1/2} a_s^{11/2}}\left[1+\frac{51e_s^2}{4}\right] + \frac{2a_s^{1/2}\Gamma_s}{M_s(GM_p)^{1/2}} - \frac{21k_2^s n_s M r_s^5}{Q_s M_s a_s^4}e_s^2 \quad \text{Eq.2}$$

$$\frac{de_s}{dt} = +\frac{57k_{2p}n_s M_s R_p^5}{Q_p M_p a_s^5}e_s - \frac{21k_2^s n_s M_p r_s^5}{2Q_s M_s a_s^5}e_s + F_{me} \quad \text{Eq.3}$$

where $R_P$ and $M_p$ are the planet's radius and mass respectively, and $M_s$, $r_s$ and $n_s$ are the satellite's mass, radius and orbital frequency, respectively. Terms with $k_2^s/Q_s$ account for dissipation inside the satellites. $F_{me}$ accounts for mutual satellite perturbations and is discussed below. Saturn's "quality factor" $Q_p$ is canonically thought to be ≥ 18,000 (assuming $k_{2p}$ = 0.341, see Goldreich and Soter 1966, Sinclair 1983). However, as we discussed in section 1, this value is very poorly constrained and a recent study (Lainey et al., 2010) suggests $Q_p$~1680, implying a much faster evolution pace. We will consider both values for $Q_p$.

For the time being no N-body code is able to handle mutual interactions between 10 to 20 bodies for 10$^{10}$ to 10$^{12}$ orbits as required here. As a consequence mutual interactions between satellites are considered in a simplified way, in the form of a forcing term (designated as $F_{me}$ in Eq. 3) that increases the magnitude of eccentricities. In Charnoz et al. (2010), $F_{me}$ was treated by using an instantaneous-encounter formalism (Greenberg et al., 1978). This technique is numerically stable for very small moons (as those considered in Charnoz et al., 2010), but leads to fast instabilities for large satellites. So the previous method is not applicable for the present study. Thus we designed a simple "toy model" for the satellites' eccentricity evolution. The main idea is the following: the eccentricities of two perturbing moons increase, on average, by an amount $\sim \Delta v/V_k$ (where $\Delta v$ is the velocity perturbation during an encounter and $V_k$ is the orbital Keplerian velocity) on a timescale comparable to the mutual synodic period of the two moons. Let $e_i$ be the eccentricity of satellite i and $T_{ij}$ their synodic period. Thus our "toy model" is as follows:

$$F_{me} = \sum_{i \neq j} \frac{(e_{fij} - e_i)}{T_{ij}}$$

**Eq. 4**

where $e_{fij}$ is the average eccentricity forced by mutual encounters between satellites i and j, defined as follows:

$$e_{fij} = Max\left\{e_i, \frac{\Delta V_{ij}}{V_{ki}}\right\}$$

**Eq. 5**

where $V_{ki}$ is the Keplerian velocity of satellite i. $\Delta V_{ij}$ represents the velocity kick induced during one conjunction between satellites i and j, and is computed using the impulse approximation (Greenberg et al., 1978). We force $e_{fij}$ to be always larger than $e_i$ so that eccentricity cannot be damped during a mutual encounter. This crude mathematical model leads to an exponential relaxation of eccentricities toward an equilibrium value on a synodic timescale and is numerically stable. It gives order of magnitudes for today's eccentricities of satellites out of mean-motion resonances (using current positions and semi-major axes of Enceladus, Tethys and Dione and Rhea), all in the range of 0.001 to a few 0.01.

We are fully aware that neglecting resonant interactions may lead to large errors in the satellites' orbital evolution (unaccounted resonant trapping, eccentricity pumping, secular effects etc..). However we consider the current work as a first attempt to understand the long-term evolution of Saturn's system including moon-ring-interactions.

## 3. Moon formation from a pure icy ring

### 3.1 Early evolution
In this section we investigate the possibility of accreting large moons (from 100 to 1000 km radius) in an initially massive disk. For now we aim to examine whether we can understand and reproduce a simple and robust observed trend: the current mass vs. distance relation of Mimas, Enceladus, Tethys, Dione and Rhea, as displayed in Fig. 1. This will be our "guideline" throughout this section. For the moment, we assume that the totality of material is made of ice. The case of a mixture of ice and silicates is treated in section 4.

All our simulations start with a narrow isolated ring, whose initial mass, $M_i$, ranges between 1 and 10 Rhea masses (where Rhea's mass $M_R = 2.3 \times 10^{21}$ kg). As a typical example, the time evolution of the system for $M_i = 4\ M_R$ is shown in Fig. 3 for three cases:

- CASE A (Fig.3.a and 3.b): $Q_p = 18,000$ and no dissipation inside the satellites
- CASE B (Fig.3.c and 3.d): $Q_p = 1680$ and no dissipation inside the satellites
- CASE C (Fig.3.e and 3.f): $Q_p = 1680$ with constant dissipation inside the satellites ($k_2^s/Q_2^s = 0.01$ constant for all satellites)

In all three cases, the ring starts spreading freely and a population of satellites appears at the disk's outer edge (i.e., Saturn's Roche limit). They grow due to mutual encounters, similarly to the small satellite formation model presented in Charnoz et al. (2010). In just $10^5$ years, moons with masses ~ $10^{17}$ kg up to ~$8 \times 10^{21}$ kg are formed with the largest gathering about 80% of the disk's initial mass (see Fig.3 and Fig.6). Subsequent accretion modifies the mass distribution while keeping almost unchanged the mass of the largest moon, $M_{big}$, which scales roughly linearly with $M_i$, as indicated in the semi-empirical formula derived from equilibrating the ring's viscous torque with the satellite's gravitational torque (see Charnoz et al., 2010 supplementary information) :

$$M_{big} = \frac{\alpha M_i}{S + 2\alpha} \qquad \text{Eq. 6}$$

and with $\alpha \sim 0.1 \times (M_p/m) \times (k/(R_L^2 \Omega_L))^{1/2}$, where m is the order of the strongest resonance (m~2 here) and $\Omega_L$ the Keplerian frequency at the distance of the Roche limit. S is the disk's surface and k is a numerical factor coming from the viscosity model (so that the viscosity is given by $\nu \sim k\sigma^2$ where $\sigma$ is the disk's surface density, see Daisaka et al., 2001), $k \sim 8 \times 10^{-8}$ $m^6 kg^2 s^{-1}$ here (see supplementary information#4 in Charnoz et al., 2010). From Eq.6, a 4-$M_R$ disk extending from Saturn's equatorial radius to Saturn's Roche limit would give birth to a moon with $M_{big} \sim 0.8$ $M_R$. However, simulations show that $M_{big} \sim 3.2 M_R$. Whereas our analytic estimate is smaller, the order of magnitude is correct. The discrepancy with simulations may be a consequence of accretion of the moons by each other, which is not considered in Eq. 6.

**3.2 Need for intense dissipation in Saturn**
In case A (with $Q_p$=18,000, see Figs. 3.a and 3.b, Case A) the masses vs. distance relation, $M(a_s)$, poorly reproduces the current configuration (large difference between the black dots and grey squares in Fig.3). Only a single large moon ends beyond 250,000 km from Saturn, instead of the three moons currently in that situation. Varying the initial conditions ($M_i$ and the initial ring location) does not provide any better match. This is not surprising because in contrast to the assumptions chosen by Goldreich and Soter (1966) the moons in our simulation (i) have not reached their final mass at the time emerge from the rings, (ii) have not formed at the location Saturn's synchronous orbit and (iii) are formed at different epoch due to the progressive spreading of the disk. Conversely, for $Q_p$=1680 (Case B) the resulting $M(a_s)$ function shares striking resemblances with the current distribution and qualitatively reproduces the current mass-distance distribution (Figures 3.b to 3.f) after 2.5 Gyr of evolution. In the resulting orbital architecture, two distinct regions are discernable. Below 220,000 km satellites interact strongly with the rings and experience a rapid outward migration. This is the "ring-controlled regime." Beyond 220,000 km, lies the "Saturn-controlled regime" in which satellite migration is slower, primarily driven by Saturn's tides. We note that the satellite systems after 2.5 Gyr and 4.5 Gyr of evolution are almost similar (compare Fig 3.c and 3.d) apart from the final location of the most massive satellite, which is highly sensitive to its mass (Eq.2). We find that in case B (in which Saturn's dissipation is included while the satellites are not dissipative, see Fig.3c and 3d) that satellites in the Saturn-controlled regime are significantly less massive compared to the current configuration. In general, the satellites end up "too far" from their host planet (either at 2.5 Gyr or 4.5 Gyr), especially those located beyond 200,000 km. A possible way to limit excessive orbital migration is to increase the value of $Q_p$. However, after performing multiple tests, increasing the value of $Q_p$ does not seem do a good job: instead it results in decreasing the number of satellites formed beyond 220,000 km: the slower the migration, the more effective the accretion process. To solve this problem we have considered dissipation inside the satellites themselves in addition to a very dissipative Saturn ($Q_p$=1680, Case C, see Fig. 3.e and Fig. 3.f). In this case the mass vs. distance distribution shows a much better agreement with the observations. Of course, that process is time-dependent and a function of the satellites interiors and compositions. By lack of constraints, we assume that $k_2^s/Q_2^s$ is constant. The large value adopted here ($k_2^s/Q_2^s = 0.01$) is an extreme case. For lower values of $k_2^s/Q_2^s$ the fit to the current satellites population gets worse, though.

Despite the simplicity of our model, the strong similarities we obtain with the current orbital configuration of Saturn's inner moons suggest that the latter is the product of strong Saturn's tides ($Q_p \sim 1680$) combined with strong dissipation within the satellites ($k_2^s/Q_2^s \sim 0.01$ here); the satellite system progressively accreted at the edge of a massive ring, spreading beyond Saturn's Roche limit. We assume that $Q_p$ is constant over the age of the Solar System which adds another degree of uncertainty to our simulations. As a consequence of all these uncertainties, it is not possible to accurately date Saturn's inner satellite system. However, we can certainly note that our simulations are compatible with formation of the most massive satellite, concurrent with the rings, at least 2.5 Gyr ago. This is compatible with both the birth of the Solar System (4.5-4.6 Gyr ago) and the date of the putative Late Heavy Bombardment (3.8-3.9 Gyr ago; Cohen et al., 2000, Tsiganis et al., 2005).

In conclusion, with a high tidal dissipation inside Saturn ($Q_p \sim 1680$ as suggested by Lainey et al., 2010), and with dissipation inside the satellites, we find that the formation of all the satellites of Saturn until (and including) Rhea may be well explained by the spreading of massive rings beyond the Roche limit.

### 3.3 Early high eccentricity episodes

A notable aspect of our simulations is the likely occurrence of high eccentricity episodes early in the satellite's histories (Fig.4). Since moons are formed in a compact orbital configuration (they all first accrete at the Roche limit, but at different times), their mutual perturbations induce high eccentricities (larger than 0.005 for the most massive, for example). Some eccentricity "peaks" are also expected (Fig.4). They result from close encounters between proto-satellites followed by sudden eccentricity decreases as a consequence of enhanced dissipation following merging (due to the enhanced dissipation of larger bodies visible in the second term in Eq. 3, which is proportional to $r_s^5/m_s$, i.e., to $r_s^2$). These early episodes of large eccentricity for the most massive satellites (Fig.4) may have had consequences for their geological activity. Positive feedback between melt generation and enhanced tidal dissipation may explain the signs of geological activity visible on some of the most massive of Saturn's mid-sized satellites (Porco et al., 2006, Stephan et al., 2010, Wagner et al., 2010) This prospect bears interesting implications that need to be studied through coupled geophysical and dynamical evolution models, which are however beyond the scope of this paper.

The mass distribution of satellites accreted in the simulations shares similarities with the observed one (Fig. 5). Apart from the smallest satellite (Atlas, mass$\sim 10^{16}$ kg, Spitale et al. 2006) the mass distribution we obtain — either with or without satellite dissipation — are somewhat similar to the current distribution (diamonds). A detailed check shows that the slope yielded by our calculation is systematically steeper than the observed one. Interestingly, there is a knee in the observed distribution at about $10^{19}$ kg that we also obtain in our simulations. The knee may be related to accretion processes occurring beyond 220,000 km and the "traffic jam" effect at this location (see section 3.4). We note that while Saturn has four satellites interior to Titan's orbit with masses $> 10^{20}$ kg, in our simulations only three such moons are formed in general. Although we do not strictly reproduce the current distribution, similarities in the slope and the presence of a knee are encouraging.

### 3.4 Accretion in several steps: the planetocentric impactors and the satellites' age

The accretion process does develop at a uniform pace and shows a clear dependence on distance to Saturn (Fig. 6). During their outward migration, satellites slow down around 220,000 km (Fig. 6.a) where ring-satellite interactions cease (more precisely when the satellite's 2:1 inner Lindblad resonance leaves the disk). As a consequence of satellites slowing down their orbital expansion when they reach this distance, they trigger a sort of 'traffic jam' effect. All satellites massive enough to migrate beyond 220,000 km finish their accretion below 250,000 km (Figs. 6.a, 6.b). So the region extending from ~200,000 km to ~250,000 km is the nursery of the most massive satellites (i.e., those larger than Enceladus). There, intense accretion occurs and may produce debris. Satellite's eccentricities at the time of accretion are in the range $10^{-3}$ to $10^{-2}$ (Fig. 4), resulting in encounter velocities in the range of only 20-200 m/s. These values are so low that the effective impact velocities at the satellites' surface will be dominated by the escape velocities of the two colliding bodies (see Eq. 19.7 of Dones et al., 2009), ranging from ~600 m/s for Rhea-sized bodies down to ~170 m/s for Mimas-sized bodies. The debris resulting from these mergers may be a source of planetocentric impactors hitting the satellites' surfaces at low velocity. This outcome of our model may give support to the hypothesis that a population of planetocentric impactors hit Saturn's icy satellites with moderate impact velocities in the past, as suggested by the icy moons' craters records (see, e.g., Smith et al., 1982; Zahnle et al., 2003; Dones et al., 2009). We propose that this population would be simply the fragments of an intense post-accretional phase for those satellites finishing their accretion beyond ~200,000 km from Saturn's centre.
In addition, differential orbital migration due to Saturn's tides favors impacts between objects of similar sizes, thus naturally promoting the formation of large basins on Saturn's mid-sized moons, in qualitative agreements with observations (like Herschel, Odysseus or Tirawa basins on Mimas, Tethys and Rhea respectively). We note however that it is not the sole scenario possible for the origin of the planetocentric population of impactors. Mosqueira and Estrada (2003b) suggest that satellitesimals may have impacted the mid-sized satellites just after their accretion, if they formed in Saturn's nebula.

The satellites' ages are plotted in Fig. 7. Age is (arbitrarily) defined as the time elapsed since the satellite reached 80% of its final mass. The oldest satellites (at about 530,000 km from Saturn) are assumed to be 2.5 Gyr old, since at this time in the simulation the satellite distribution matches pretty well the current distribution (see Fig. 3.e). Fig.7 shows that satellites beyond 220,000 km have similar ages and are of the same generation, while the moons below 200,000 km are at least 1 Gyr younger. This age difference between the Mimas-like moons and the Rhea-like moons is systematically observed in our simulations.

The systematic age trend observed in our results (the age is an increasing function of mass) may be qualitatively explained on simple arguments: when the ring is young, its surface density is high. Then, the flux of material through the Roche Limit is also high. Consequently massive satellites are formed early in the history of the system (see Figure 3). In addition, the more massive a satellite, the faster its orbital expansion (due to Saturn's tides). Hence large satellites will efficiently sweep-up neighbouring satellitesimals on their way outward, and grow even more rapidly. Conversely, later in the history of the system, when the ring has significantly spread (through the Roche Limit and onto the planet's atmosphere) the material flux at the Roche Limit is weaker due to the ring's lower surface density. Consequently, smaller satellites are formed that will experience a slow orbital expansion. As a direct consequence, small satellites are systematically younger and closer to the planet than large satellites, as observed in Figure 6.

These results imply that Mimas may have accreted after the putative LHB bombardment (about 3.8 Gyr ago). Indeed, The present scenario offers an explanation for the existence of an object like Mimas, which was likely to be disrupted during the LHB (Charnoz et al., 2009).

Re-accreation of this object at a temperature of 70 K would have likely led to a rubble-pile (e.g., as an analog, 24 Themis is an example of rubble-pile reaccreted from a larger parent body in conditions where planetary materials cannot easily flow and promote shape relaxation, Castillo-Rogez and Schmidt 2010). Quite the opposite, Schenk (2011) pointed out that Mimas is surprisingly spheroidal for an old satellite subject to a very long impacting history involving large interlopers (this author takes Iapetus as a reference for the type of heavy cratering that one would expect to find at Mimas). Thus Mimas' very existence, as well as its peculiar cratering properties not easily explained in the context of the LHB, may be indirect evidence for a recent time of formation. In this context Mimas' craters may be largely due to planetocentric impactors rather than heliocentric ones, including the Herschel basin.

The inconsistency of the cratering records between Iapetus and Rhea (Smith et al., 1981; Lissauer et al., 1988; Dones et al., 2009) and Rhea's apparent youth, might be understood if Iapetus formed jointly with Saturn while Rhea formed latter, for instance 2.5 Gyr ago. However, our simplified dynamical model does not allow us to construct an accurate chronology.

### 3.4 What about the rings?

The different panels of Fig. 3 show that the final density profiles of the ring are very similar after 2.5 and 4.5 Gyr of evolution. The ring profile is almost independent from the value of $Q_s$ or $Q_p$. The ring ends up with a total mass a few times Mimas' mass, compatible with current estimates (Charnoz et al., 2009b; Robbins et al., 2010). The current surface density of Saturn's rings is thought to be ~ 400kg/m$^2$ in the A ring (beyond 127,000 km, Robbins et al., 2010) and perhaps greater in the B ring (but it is not well constrained), in qualitative agreement with our results. However, the most obvious discrepancies with today's ring are (i) the location of the inner edge, around 75,000 km today, rather than 60,000 km in our model (in contact with Saturn's atmosphere) while the densest region today is around 110,000 km (the B ring), rather than 70,000 km; and (ii) the absence of a Cassini division in our simulation. Note however that no other model provides an explanation for these two features. Salmon et al. (2010) have shown that the final rings' mass depends very weakly on the initial conditions as the ring spontaneously tends toward a state where it is marginally gravitationally instable everywhere (so that the so-called Toomre parameter, Q, is everywhere close to unity). As a direct consequence the final mass of the rings is always comparable to Mimas' mass whatever its initial mass (see Salmon et al., 2010, for detailed explanations).

### 4. Forming the current inner moon system via heterogeneous accretion

The absence of rocky material in Saturn's rings contrasts sharply with its occurrence (from ~6% to ~60%) inside the mid-sized moons (Harris, 1984, Nicholson et al., 2005, Thomas et al., 2007, Cuzzi et al., 2009, Matson et al., 2009, Thomas 2010). This may appear in contradiction for any model attempting to form Saturn's icy moons from a rings that appears devoid of rock (Poulet et al., 2003, Nicholson et al., 2005, Cuzzi et al., 2009). This discrepancy may be seen a weakness of the recent work by Canup (2010) that invokes a fully differentiated Titan sized-satellite as the ring progenitor to explain the pure composition of today's rings. However we think it is still possible to envision a scenario of moon formation from ice dominated rings.

The following describes and simulates a scenario for producing Saturn's inner moons and provides a context for explaining the variations in rock content across this system, while explaining the eventual depletion of the rings in rock. This model requires that large chunks of silicate (> 10-100 km) were present in the rings, post-formation. This could be possible, for example, if the ring's progenitor was only partially differentiated at the time of disruption, which is still a matter of debate. In this context it is possible to create today's Saturn's icy moons from the rings, with stochastic silicates abundances, while the resulting ring system is left largely devoid of silicates. The two key ingredients are :
- "heterogeneous" accretion in the rings (i.e silicate chunks accreting an icy shell)
- migration of protosatellites in the young massive rings due to tidal interactions with the disk

**4.1 Heterogeneous accretion: principles and model**
Whatever the process that destroyed the icy mantle of the ring's progenitor (either by tides at the end of Saturn's formation, as investigated in Canup 2010, or after a cometary impact during the LHB, as suggested in Charnoz et al., 2009a) some silicates may remain in the icy debris if the parent body was only partially differentiated. While silicate grains condensed in the Solar nebula are expected to be intimately mixed with ice at the microscale level, several processes can be suggested to explain the presence of large chunks of rocky material in the rings progenitor. For example, this large progenitor could have accreted in Saturn's nebula from proto-satellites a few tens km in radius subject to early ice melting and separation of a rock-rich core as a consequence of short-lived radioisotope heating, and/or tides. These objects could have then merged into a large satellite, or, as an alternative, been disrupted as a consequence of collisional grinding into rock- or ice-dominated debris. A second generation of larger satellites could have then accreted silicate fragment now several tens of kilometres large. In the Canup's model, that proposes that the ring formed just at the end of Saturn's formation, a tight timing is necessary between the formation of Titan and its differentiation, that is still a matter of debate (see e.g. Mosqeira and Estrada 2011). On the other hand, in case of full differentiation, the total silicate mass inside Saturn's inner moons represents only ~2% of the mass of Titan's silicates. So, even if the ring progenitor was fully differentiated, such a small mass of silicates embedded with ice is a plausible assumption. Note that the ring progenitor could also be a massive heliocentric interloper, but its origin would have to be determined. However, Charnoz et al., (2009) have shown that the flux of small and undifferentiated bodies (< 100 km) would overcome by far the flux of single massive bodies if this happened during the LHB. Then, the cometary shower on giant-planets would create a debris disk surrounding the planet in which ice and silicates would be intimately mixed, a situation that does not appear compatible with the formation mechanism investigated in this study (see below). So instead we assume that an almost differentiated body was destroyed inside Saturn's Roche Limit, where it left a collection of massive chunks of silicates embedded in a ring of icy debris.

What is the fate of these silicate chunks embedded in the massive icy ring?
A silicate chard is resistant to tidal shattering beyond 90,000 km from Saturn whereas the tidal shattering of ice occurs at about ~130,000 km. Thus the location of the Roche Limits $R_L$ depends on the material density $\rho_M$ through $R_L \sim 2.456\, R_p (\rho_p / \rho_M)^{1/3}$ giving $R_{L,I} \sim 136,000$ km for ice and $R_{L,S} \sim 90,000$ km for silicates.

As a direct consequence, between 90,000 km and 136,000 km from Saturn, rock boulders can accrete and gather the surrounding icy material, whereas the accretion of icy debris is impeded by tidal splitting. As a direct consequence of three-body orbital dynamics, the icy material coming from the icy debris disks flows around the silicate chunk through its Lagrange points L1 and L2. The same mechanism is responsible for the formation of Pan's equatorial ridge (Charnoz et al., 2007). In this context, a rock shard will progressively fill its Hill sphere with ice. This is somewhat the inverse process as the peeling of a differentiated satellite larger than its Hill sphere, as simulated by Canup (2010). A Hill sphere is the region of gravitational reach of a satellite in the potential field of its planets. It is approximately a tri-axial ellipsoid with axes ratio 3:2:2 (Porco et al., 2007) centred on the body (Fig. 8). The long axis joins the Lagrange points L1 and L2, and its radius is

$R_h = a_s (M_s/3M_p)^{1/3}$ **Eq. 7**

The volume of the Hill sphere $V_h$ is (Porco et al., 2007)

$$V_h \approx \frac{2\pi \ln(2+\sqrt{3})}{3\sqrt{3}} R_h^3$$

**Eq. 8**

So substituing Eq.7 into Eq.8 simply gives:

$$V_h = \eta a_s^3 \left(\frac{M_s}{M_P}\right)$$

**Eq. 9**

With $\eta = 2\pi \ln(2+\sqrt{3})/(9\sqrt{3}) \approx 0.53$. A fraction of the Hill Sphere is occupied by the rocky chunks while the empty space will be progressively filled in with icy debris, that will, in turn, increase further the volume of the Hill Sphere (because they increase the body's mass). The icy particles accreted at the surface of the chunk will form a random accumulation of ice, with a high degree of porosity (Porco et al., 2007). Let $\rho_e$ represent the effective density of the ice accumulated at the surface of the silicate chunks, and let $m_s$ and $m_i$ be the total mass of silicate and ice in the body, respectively. Since $M_s = m_s + m_i$ and the total volume is $V = m_s/\rho_s + m_i/\rho_e$ it is easy to compute the asymptotic ice to silicate mass ratio ($m_i/m_s$) of a protosatellite accreted at a distance $a_s$ on a rocky chunk of mass $m_s$ by solving $V = V_h$. We find:

$$\frac{m_i}{m_s} = -\frac{\rho_e}{\rho_s}\left(\frac{M_p - \eta \rho_s a_s^3}{M_p - \eta \rho_e a_s^3}\right)$$

**Eq. 10**

Since the rocky chunk is not filling its Hill sphere we have $M_p - \eta \rho_s a_s^3 < 0$. Thus, from Eq. 10 we see that $m_i/m_s$ is positive if the ice effective density ($\rho_e$) is smaller than a critical density $\rho_c = M_p/\eta a_s^3$. If $\rho_e < \rho_c$ then the icy material flowing on the silicate chunk is underdense and will ultimately fill the body's Hill sphere. If $\rho_e > \rho_c$ the constraint $V = V_h$ has no solution with positive $m_i$ ; in fact, the Hill sphere grows rapidly as the material is accreted, so that the accretion goes on without limit. In that case, we obtain a protosatellite mostly composed of ice. Thus in order to form satellites with similar amounts of ice and silicate we must be in the regime $\rho_e < \rho_c$. For 80,000 km $< a_s <$ 140,000 km, $\rho_c$ decreases from 2100 kg/m$^3$ down to 450 kg/m$^3$. This means that if the ice accreted at the surface of the silicate chunk has a porosity greater than 60% (i.e., $\rho_e <$ 450 kg/m$^3$) then proto-satellites with similar rock-to-ice ratios (like Mimas, Enceladus, Dione or Rhea) may form everywhere in the disk, provided that a silicate shard is initially present. More precisely for $\rho_e < \rho_c$ the silicate mass fraction ($f \equiv m_s/(m_i + m_s)$) is:

$$f = \frac{\rho_s (M_p - \eta \rho_e a_s^3)}{M_p (\rho_s - \rho_e)}$$

**Eq. 11**

In Fig. 9 we plot $f$ as a function of $\rho_e$ at $a_s$=130,000 km (near the disk's outer edge). A silicate mass fraction of about 50% is obtained for $\rho_e \sim 260$ kg/m$^3$, corresponding to an ice porosity about 70%. This high porosity is comparable to the current porosity inferred for Saturn's smallest moons (Porco et al., 2007) and is also in agreement with the outcome of N-body simulations of aggregate formation in Saturn's rings at ~130000 km from Saturn (Porco et al., 2007). Hence, silicate chunks embedded in the disk may accrete an icy shell until their Hill sphere is filled with porous ice, resulting in the formation of proto-satellites with a silicate core, that is *not* necessarily spherical.

What is the fate of these proto-moons? Do they stay in the disk or are they expelled from the ring system?

**4.2 Numerical simulation of heterogeneous accretion coupled with migration**
As the proto-moons grow within the disk they undergo tidal interaction with the disk and may be expelled from it. Several regimes of migration in the disk are :

- A "Type-I-like" migration, in which a proto-moon does not open a gap: the migration regime in this case has been studied by Crida et al. (2010). In contrast to gaseous protoplanetary disks, there is no pressure buffer, so the classical type I migration isn't at play. In consequence , in Saturn's rings, the motion is directed in the local direction of decreasing density, because low density regions exert a weaker, on the satellite, than high density regions (MeyerVernet & Sicardy 1987).
-  A Type-II migration, in which a proto-moon opens a gap. Its direction if motion is then controlled by the local viscous flow of disk material. The moon is viscouly "locked" in the disk, similarly to standard type II migration in protoplanetary disks (Lin and Papaloizou 1986)

Note that in our numerical model (see section 2) we do not use any prescription either for type-I or type-II migration. These two regimes naturally appear as a consequence of our explicitly taking into account the satellites' torques onto the disk and the disk's viscous torque (see section 2 or Charnoz et al., 2010 and Salmon et al., 2010). The two regimes are the following: when a satellite's gravitational torque is stronger than the disk's viscous torque, a gap opens around the satellite (see e.g. Fig. 10) and the satellite is locked with the disk's viscous evolution (type II migration). Otherwise if a satellite is not able to open a gap, it still moves relatively to the disks because of the disk's gravitational torque (type-I like migration).

We have computed the migration of a proto-moon in the disk and the action it exerts in return on the disk. In addition, we have coupled this orbital evolution with the physics of heterogeneous accretion (see section 4.1), so that we can compute the position as well as the silicate fraction of a proto-moon at each time step. The algorithm is described below.

**4.2.1 Heterogeneous accretion algorithm**
An heterogeneous proto-moon contains a central silicate core with mass $m_s$ and density $\rho_s$=3000 kg/m$^3$ (averaged between a pure hydrated silicate and anhydrous silicate material) and an outer shell of porous ice with mass $m_i$ and effective density $\rho_e$ arbitrarily chosen as 450

kg/m$^3$ (equivalent to a porosity of about 60%). A body at location $a_s$ and mass $m=m_s+m_i$ can accrete, at most, all the disk material located in a narrow ring extending radially from $a_s$-$R_h$ to $a_s$+$R_h$. This mass is called $M_{max}$. Because of tidal forces acting on the proto-moon, $M_{max}$ may actually not be entirely accreted, and the condition for accretion is as follows: the material accreted by the proto-moon must be contained between the volume of the body's Hill sphere $V_h$ and the physical volume of the moon V (Canup and Esposito 1995). So the mass of icy material exchanged with the disk is $M_{exchanged}$=min{$M_{max}$ , $\rho_e$ ($V_h$-V)}. If $M_{exchanged}$>0 then there is a net accretion of ice, and $m_i$ increases by the amount $M_{exchanged}$, while the disk surface density decreases accordingly. If $M_{exchanged}$ < 0 tides shatter the icy shell and the icy material is fed back to the disk. Consequently, $m_i$ decreases by an amount ‖$M_{exchanged}$‖, and the local disk surface density increases accordingly.

### 4.2.2 Results: Forming moons with varying silicate contents

Figure 10 displays the moons masses (right ordinate) and the disk surface density (left ordinate). The symbols are color-coded as a function of the silicate mass fraction in the moons. We start with a pure silicate chunk of mass 2×10$^{21}$ kg (about Rhea's mass) in a disk of pure icy material of about 4 Rhea masses. The chunk accretes an icy shell on its surface (section 4.1 and Fig. 11) while migrating outward. It is massive enough to open a gap (Fig. 10) and so undergoes type II migration. It is ejected from the disk in a couple of years. While the moon starts as a pure silicate body, it accretes a shell of ice representing about 25% of its total mass (Fig.11) so that its silicate mass fraction drops to ~65%. While the moon migrates outward its ice content slowly increases. As the moon is ejected from the disk, it merges with another moon made of pure ice that formed at the ring's outer edge. After the collision, the icy content of the larger moon jumps (Fig.11). This decreases the silicate fraction of the moons further, down to about 40% (Fig.10 and Fig.11), which now approaches the current composition of Mimas, Enceladus, Dione and Rhea. On longer timescales, this moon will migrate outward while additional pure-ice moons will be assembled at the disk's outer edge. These moons mostly made up of ice are similar to Tethys (which contains from 7% to 10% of silicate, depending on the porosity). Unfortunately, since computing the evolution of a moon embedded in the rings is very numerically intensive, we could not model the evolution of a disk populated with several moonlets, over the age of the Solar System.

### 4.2.3 Depleting the ring's silicate content while forming Pan, Daphnis and the propellers

Is the above example representative of the fate of any silicate chunks starting in the primordial icy disk? We have verified numerically (Fig. 12) that any silicate body larger than 10 km (~10$^{15}$ kg) is rapidly tidally expelled from the disk in the direction of the local downward surface-density gradient. Bodies Smaller that 10$^{15}$kg could not be considered because of computer limitations. The ejection timescale depends on the silicate chunk's mass, its starting position and the initial density distribution of the ring. These three parameters determine whether the moon is evolving through type I or type II migration. In addition, since the disk's surface density decreases with time, it is possible to switch from Type I to Type II migration (because the viscosity is an increasing function of density, Daisaka et al., 2001). Some moons with masses greater than 10$^{17}$ kg are ejected in less than 10$^5$ years (Fig. 12) while lighter moons can take up to 10$^8$ years to leave the rings depending on their initial positions and the disk surface density. Whatever the migration regime the end result is always an ejection from the disk as illustrated in Fig.12. Most remarkably, a recent analysis of Pan and Daphnis (two moons that have opened gaps in their orbits in the Encke gap and Keeler gap, respectively) has concluded that each may hide a central dense shard surrounded by an underdense icy shell (Porco et al. 2007). This result concurs with the scenario proposed in this study: Pan and Daphnis could hide in their centres a chunk of silicate, remnant from the ring's progenitor, below an ice shell. It may be possible that Saturn's rings' buried moonlets (called "propellers", see Tiscareno et al., 2006 and Sremčević et al., 2007) may share a similar origin: small pieces of silicates, coming from the ring progenitor, covered with ice that failed to migrate out from the ring system because they are not massive enough for their migration to be fast enough. Actually, the migration of such small bodies may not be dominated by type I like regime that we observed here (see Crida et al. 2010 and Rein & Papaloizou 2010 for an analysis of the migration of the "propellers") but may be subject to some stochastic (and so very innefficient) migration process due to random encounters with nearby gravity wakes.

## 5. Conclusion : Summary of main results

In the present paper we have identified and modelled the mechanisms driving the accretion of proto-moons at the edge of an initially massive saturnian ring system initially embedded in Saturn's Roche Limit. Our main conclusions are

- It is possible to form all Saturn's mid-sized moons, from Mimas to (and including) Rhea, provided that the primordial disk was a few times Rhea's mass.
- Migration of these moons toward their current locations requires that Saturn's average dissipation coefficient, $Q_p$, is less than 2000. This is compatible with the value of 1680 recently suggested by Lainey et al. (2010) based on satellite's ephemeris. $Q_p \sim 18000$ is generally considered as a minimum value within to the four classical assumptions of the Goldreich & Soter work (Goldreich & Soter 1966, see discussion in section 1). However, in the present work we find that (i) the satellites' masses progressively increase with time (ii) the satellites are formed at different epochs (iii) they appear at the Roche Limit, and not at the synchronous orbit (whereas this difference may play a minor role). This departure from the framework assumed by Goldreich and Soter explains why our dynamical model requires a $Q_p$ much lower than 18000 in order to implant the satellites close to their current location.
- Our results suggest a new chronology for the ages of the satellites. In that framework, a Mimas-like satellite could be about 1 to 1.5 Gy younger than a Rhea-like satellite.
- The satellites' formation is tied to the ring's formation, which could have occurred anytime between 2.5 Gy and 4.5 Gy old, contemporaneously with Saturn's formation or as a product of the Late Heavy Bombardment some ~3.8 Gyr ago.
- Episodes of large eccentricity increase might have triggered intense tidal-heating periods in the newly-formed satellites, which might explain the visible signs of past geological activity at the surface of Saturn's satellites.

If the ring progenitor was only partially differentiated then it is possible to simultaneously explain (a) the diversity of silicate abundances among Saturn's mid-sized moons and (b) a silicate-free ring system, and potentially, the origin of Pan, Daphnis and the propellers. The key idea is that within Saturn's Roche limit silicate material is dense enough to accrete its surrounding material, while icy material cannot accrete:
- If chunks of silicate > 100 km are initially embedded in the primordial massive disk they can rapidly accrete a shell of icy material inside their Hill sphere, leading to the formation of pre-differentiated proto-moons.
- These proto-moons are then expelled from disk due to gravitational interaction with the ring. Later on, they can coalesce and merge their content.
- This process exports the silicates from the ring to the satellites, leading to the progressive purging of the ring from its rock.

- In this context, the final silicate mass fraction of these moons are the result of stochastic accretion and of the stochastic initial distribution of silicate chunks. This process may result either in moons with high silicate fraction (about 50%), or down to 0% (Fig. 10), in qualitative agreement with today's configuration (Fig.2). A moon like Tethys, dominated by water ice, may simply indicate that the icy clumps it accreted from never encountered a silicate chunk.
- Depending on the long-term thermal evolution of satellites interiors, the rocky chunks may remain irregular, potentially explaining the departure from hydrostatic equilibrium of Enceladus' shape (Thomas et al. 2007), as explicitly suggested earlier by Schenk and McKinnon (2008). This scenario can also explain the discrepancy between global shape and gravity data as observed on Rhea (Nimmo et al., 2010)

This model unifies under the same framework Saturn's rings and Saturn's mid-sized moons. Moons currently orbiting below Saturn's Roche Limit (like Pan or Atlas and the propellers) are interpreted as remnants of Saturn's rings formation, compatible with a dense core of silicate coated with ice (consistently with Porco et al., 2007). Satellites beyond Saturn's Roche Limit may have formed jointly with, or after Saturn's rings. During this process, the rings progressively lost their silicate content explaining why the youngest moons (those close to Saturn's rings) are ice rich.

Strictly speaking the mechanism described here does not necessarily imply that the ring progenitor was a large Saturn's satellite shattered by tides (as in Canup 2010) however it is compatible with it. It could be also a heliocentric interloper massive enough to have reached a close-to-complete differentiation, with a mass of about few Rhea's mass that passed close to Saturn and was tidally shattered. It could also be a large satellite destroyed by a cometary impact as suggested in Charnoz et al. (2009). However the probabilities of these last two possibilities still have to be quantified.

This model complements and extends the work of Charnoz et al. (2010) about the origin of the small moons while being consistent with the recent suggestion that Saturn's rings were formed from the tidal splitting of a Titan-sized moon, and were therefore initially very massive, which is in agreement with their viscous evolution over a few billions years (Salmon et al. 2010). However we emphasize that our scenario is somewhat independent from the rings formation timeframe and would apply as well if these formed during the LHB. While the current work cannot discriminate between the two timeframes, it may be possible to address this question in the future from satellites' geological evolution and from the value of Saturn's $Q_p$. On the one hand, this work opens the door to new research on the geophysical evolution of Saturn's satellites for initial conditions very different from those considered in the classical formation scenario. On the other hand an independent determination of Saturn's $Q_p$ based on observational arguments, would prove a critical test for the scenario presented in the current paper that requires a low value (typically under 2000) to implant the moons at their current locations, either the mid-sized moons formed 4.5 Gyr ago or later, during the LHB.

This work emphasizes that planetary moons and rings are two interconnected and genetically linked components of a fundamentally unique system.

**ACKNOWLEDGEMENT**
We thank warmly Paul Estrada for his detailed review that helped us to increase the quality of the paper. Part of this work was carried out at Université Paris Diderot with a partial funding from a CAMPUS SPATIAL grant, as well as by Institut Universitaire de France (IUF). It was also supported by CEA/IRFU/SAp, and by an EMERGENCE UPMC grant (contract number EME0911) and by the Cassini project. Part of this work has been carried out at the Jet Propulsion Laboratory, California Institute of Technology, under contract to NASA. US Government sponsorship acknowledged. Support from the French Programme National de Planétologie (PNP) is also acknowledged.

**REFERENCES**

Barr A.C., Canup R. M., 2008. Constraints on gas giant satellite formation from the interior states of partially differentiated satellites. *Icarus* **198**, *163-177*

Benz W., Asphaug E., 1999. Catastrophic disruptions revisited. *Icarus* **142**, 5–20.

Canup, R. M., Esposito, L. W., 1995.Accretion in the Roche zone: Coexistence of rings and ring moons. *Icarus* **113**, 331-352

Canup, R.M., 2004. Simulation of a late lunar-forming impact. *Icarus* **168**, 433-456

Canup R.M., Ward W.R., 2006. A common mass scaling for satellite systems of gaseous planets. *Nature* **411**, 834-839

Canup R., 2010. Origin Of Saturn's Rings and inner moons by mass removal from a lost Titan-sized satellites. *Nature* **468**, 943-946

Castillo-Rogez J., D. L. Matson, C. Sotin, T. V. Johnson, J. I. Lunine, P. C. Thomas, 2007. Iapetus' Geophysics: Rotation Rate, Shape, and Equatorial Ridge, Icarus, doi:10.1016/j.icarus.2007.02.018.

Castillo-Rogez, J. C., Schmidt, B. E. (2010), Geophysical evolution of the Themis family parent body, Geophys. Res. Lett., 37, L10202, doi:10.1029/2009GL042353

Charnoz S., Brahic A., Thomas P., Porco C., 2007. The equatorial ridges of Pan and Atlas: Late accretionary ornaments? *Science* **318**, 1622-1624

Charnoz S., Morbidelli A., Dones J., Salmon A., 2009a. Did Saturn's rings form during the Late Heavy Bombardment? *Icarus* **199**, 413-428

Charnoz S., Dones L., Esposito L.W., Estrada P.R., Hedman M.M. 2009b. Origin and evolution of Saturn's ring system. In *Saturn from Cassini-Huygens*, M. K. Dougherty, L. W. Esposito, T. Krimigis, Eds. (Springer Netherlands), pp. 535-573

Charnoz S., Salmon J., Crida A., 2010. The recent formation of Saturn's moonlets from viscous spreading of the main rings. *Nature* **465**, 752-754

Cohen B.A., Swindle T.D., Kring D.A., 2000. Support for the lunar cataclysm hypothesis from lunar meteorite impact melt ages. *Science* **290**, 1754

Crida A., Papaloizou J. C. B., Rein H., Charnoz S., Salmon J., 2010. Migration of a Moonlet in a Ring of Solid Particles: Theory and Application to Saturn's Propellers. AJ **140**, 944-953

Cuzzi, J.N.; Estrada, P.R., 1998. Compositional evolution of Saturn's rings due to meteoroid bombardment. *Icarus* **132**, 1-35.

Cuzzi J., Clark R., Filacchione G., French R., Johnson R., Marouf E., Spilker L., 2009. Ring particle composition and size distribution, in *Saturn from Cassini-Huygens*, M. K. Dougherty, L. W. Esposito, T. Krimigis, Eds. (Springer Netherlands), pp. 535-573.

Daisaka H., Tanaka H., Ida S., 2001. Viscosity in dense planetary rings with self-gravitating particles. *Icarus* **154**, 296-312.

Dermott S.F., Malhotra R., Murray C.D., 1988. Dynamics of the Uranian and Saturnian satellite systems - A chaotic route to melting Miranda? *Icarus* **76**, 295-334.

Dones, L., Chapman C.R., McKinnon W.B., Melosh H.J., Kirchoff M.R., Neukum G., Zahnle K.J, 2009. Icy satellites of Saturn: impact cratering and age determination. in *Saturn from Cassini-Huygens*, M. K. Dougherty, L. W. Esposito, T. Krimigis, Eds. (Springer Netherlands), pp. 613-635

Estrada P.R., Mosqueira I, 2006. A gas-poor planetesimal capture model for the formation of giant planet' satellite systems. *Icarus* **181**, 486-509

Eluszkiewicz, J., Leliwa-Kopystynski, J. 1989. Compression effects in rock-ice mixtures: an application to the study of satellites, Phys. Earth Planet. Int. 55, 387-398.

Johnson T.V., Estrada P.R., 2009. Origin of the Saturn System. In *Saturn from Cassini-Huygens*, M. K. Dougherty, L. W. Esposito, T. Krimigis, Eds. (Springer Netherlands), pp. 535-573

Estrada, P. R., Mosqueira, I., Lissauer, J. J., D'Angelo, G., and Cruikshank, D. P. 2010. Formation of Jupiter and conditions for the accretion of the Galilean satellites. Europa (McKinnon, W., Pappalardo, R., Khurana, K., eds.), University of Arizona Press, Tucson

Goldreich P., 1965. An explanation of the frequent occurrence of commensurable mean motions in the Solar System. *MNRAS* **130**, 159-181

Goldreich P., Soter S. 1966. Q in the Solar System. *Icarus* **5**, 375-389

Gomes R., Levison H.F., Tsiganis K., Morbidelli A., 2005. Origin of the cataclysmic Late Heavy Bombardment period of the terrestrial planets. *Nature* **435**, 466-469.

Harris A., 1984. The origin and evolution of planetary rings. In *Planetary Rings*, R. Greenberg, A. Brahic, Eds., (Univ. Arizona Press) pp. 641-659.

Gavrilov S.V., Zharkov Z.N., 1977. Love numbers of the giant planets. *Icarus* **32**, 443-449

Greenberg R., Hartmann W.K., Chapman C.R., Wacker J.F. Planetesimals to planets. Numerical simulation of collisional evolution. *Icarus* **35**, 1-26 (1978)

Kaula W.M., 1965. Tidal dissipation by solid friction and the resulting orbital evolution. *Rev. of Geophys. and Space Phys*, **2**, 661-685

Hartmann, W.K. 1984. Does crater 'saturation equilibrium' occur in the solar system? *Icarus* **60**, 56-74.

Iess,L., Rappaport, N.J.,Jacobson, R.A.,Racioppa, P.,Stevenson, D.J.,Tortora, P.,Armstrong, J.W., Asmar,S.W., 2010. Gravity Field, Shape, and Moment of Inertia of Titan. *Science* **327**,1367

Johnson, Torrence V., Lunine, Jonathan I., 2005. Saturn's moon Phoebe as a captured body from the outer Solar System, *Nature* **435**, 69-71

Johnson T.V., Estrada P.R. 2009. Origin of the Saturn system. In *Saturn from Cassini-Huygens*, M. K. Dougherty, L. W. Esposito, T. Krimigis, Eds. (Springer Netherlands), pp. 55-74

Katrin S. et al., 2010. Dione's spectral and geological properties. *Icarus* **206**, 631-652

Kirchoff, M. R., Schenk, P. M., 2010. Impact cratering records of the mid-sized icy saturnian satellites, *Icarus* **206**, 485-497

Lainey V., Karatekin Ö, Desmars J., Charnoz S, 2010. Saturn tidal dissipation from astrometric observations. *BAAS* **41**, 936

Lin D. N. C., Papaloizou J. C. B., 1986. On the tidal interaction between protoplanets and the protoplanetary disk. III - Orbital migration of protoplanets. ApJ **309**, 846-857.

Lissauer J.J., Squyres S.W., Hartmann W.K., 1988. Bombardment history of the Saturn system. *J. Geophys. Res* **93**, 13776-13804

Matson D. L., Castillo-Rogez J. C., McKinnon W. B., Sotin C., Schubert G., The thermal evolution and internal structure of Saturn's midsize icy satellites, In: Saturn after Cassini-Huygens, Eds: R. Brown, M. Dougherty, L. Esposito, T. Krimigis, H. Waite, pp 577-612

Meyer, J., Wisdom, J. 2007. Tidal heating in Enceladus. *Icarus* **188**, 535-539

Meyer-Vernet N., Sicardy B., 1987. On the physics of resonant disk-satellite interactions. *Icarus* **69**, 182-212

Morbidelli A., Levison H.H., Tsiganis K., Gomes R., 2005. Chaotic capture of Jupiter's Trojan asteroids in the early Solar System. *Nature* **435**, 462-465

Mosqueira I., Estrada P.R., 2003a. Formation of the regular satellites of giant planets in an extended gaseous nebula II: subnebula model and accretion of satellites. *Icarus* **163**, 198-231

Mosqueira I., Estrada P.R., 2003b. Formation of the regular satellites of giant planets in an extended gaseous nebula II: satellite migration and survival. *Icarus* **163**, 232-255

Mosqueira I., Estrada P.R., Charnoz S., 2010. Deciphering the origin of the regular satellites of gaseous giants –Iapetus: The Rosetta ice-moon. *Icarus* **207**, 448-460

Mosqueira I., Estrada P.R., 2011. On the origins of the Saturnian moon-ring system.  42$^{nd}$ **LPI** conference, #2151

Murray C.D., Dermott S.F., 1999. *Solar System Dynamics* (Cambridge Univ. Press, New York).

Nicholson P.D., French R.G., Campbell D.B., Margot J-L, Nolan M.C., Black G.J., Salo H.J., 2005. Radar imaging of Saturn's rings. *Icarus* **177**, 32-62.

Nimmo, F., Bills, B. G., Thomas, P. C., Asmar, S. W. , 2010. Geophysical implications of the long-wavelength topography of Rhea. *J. Geophys. Res.*, **115**, E10008

Owen T., Coradini A., Bar-Nun A., Roush T. 1995. Why are Saturn's inner satellites so white? *Bull. Amer. Astron. Soc.* **27**, 1167.

Peale S.J., Cassen P., 1978. Contribution of tidal dissipation to lunar thermal history. *Icarus* **36**, 245-269

Porco C.C. et al. 2006. Cassini observes the active south pole of Enceladus. *Science* **311**, 1393-1401

Porco C.C., Thomas P.C., Weiss J.W., Richardson D.C., 2007. Physical characteristics of Saturn's small satellites provide clues to their origins. *Science* **318**, 1602-1607.

Poulet F., Cruikshank D.P., Cuzzi J.N., Roush T.L., French R.G., 2003. Composition of Saturn's rings A, B, and C from high resolution near-infrared spectroscopic observations. *A&A*. **412**,305–316.

Robbins S.J., Stewart G.R., Lewis M.C., Colwell J.E., Sremčević., 2010. M. Estimating the masses of Saturn's A and B rings from high-optical depth N-body simulations and stellar occultations. *Icarus* **206**, 431-445

Salmon J., Charnoz S., Crida A., 2010. Long-term and large-scale viscous evolution of dense planetary rings. *Icarus* **209**, 771-785

Sasaki T., Stewart G.R., Ida S., 2010. Origin of the different architectures of the jovian and saturnian satellites systems. *ApJ* **714**, 1052-1064

Schenk, P. M., McKinnon, W. B., 2008. The Lumpy Shape of Enceladus and Implications for the Interior, 39th Lunar and Planetary Science Conference, 2523.

Schenk, P. M., 2011. Geology of Mimas? *Lunar. Planet. Sci. Conf.* **42**, 2729.

Schenk P.M., Murphy S.W. The rayed craters of Saturn's icy satellites (including Iapetus): Current impactor populations and origins. *Lunar. Planet. Sci. Conf.* **42**, 2098

Schubert, G.; Hussmann, H.; Lainey, V.; Matson, D. L.; McKinnon, W. B.; Sohl, F.; Sotin, C.; Tobie, G.; Turrini, D.; van Hoolst, T., 2010. Evolution of icy satellites, *Space Sci. Rev.*,**153**, 447-484

Schubert, G., Anderson, J. D., travis, B. J., Palguta, J., 2007. Enceladus: Present internal structure and differentiation by early and long-term radiogenic heating, *Icarus* **188**, 345-355

Sinclair, A.T, 198. A re-consideration of the evolution hypothesis of the origin of the resonances among Saturn's satellites. In: Dynamical Trapping and Evolution in the Solar System, Y.Y Markellos, Y.,Kozai, (Eds.), (Reidel, Dordrecht) pp 19-25

Smith B.A, et al., 1982.A new look at the Saturn system: the Voyager 2 images. *Science* **215**, 504-537

Squyres, S.W., Reynolds, R.T., Summers, A.L., Shung, F., 1988. Accretional heating of the satellites of Saturn and Uranus. J. Geophys. Res. 93, 8779.

Sremčević M., Schmidt J., Salo H., Seiß M., Spahn F., Albers N., 2007. A belt of moonlets in Saturn's A ring. *Nature* **449**, 1019-1021

Stophan et al., 2010. Dione's spectral and geological properties. *Icarus* **206**, 631-652

Thomas P.C., Burns J.A., Helfenstein P., Squyres S., Veverka J., Porco C., Turtle E.P., McEwen A., Denk T., Giese B., Roatsch T., Johnson T.V., Jacobson R.A, 2007. Shapes of Saturnian icy satellites and their significance. *Icarus* **190**, 573-584

Thomas P.C, 2010. Sizes, shapes and derived properties of the saturnian satellites after the Cassini nominal mission. *Icarus* **208**, 395-401

Tiscareno M.S., Burns J.A., Hedman M.M., Porco C.C., Weiss J.W., Dones L., Richardson D.C., Murray C.D., 2006. 100-metre-diameter moonlets in Saturn's A ring from observations of 'propeller' structures. Nature 440, 648-650

Tsiganis K., Gomes R., Morbidelli A., Levison H.F., 2005. Origin of the orbital architecture of the giant planets of the Solar System. *Nature* **435**, 459-461.

Wagner R.J., Neukum G., Giese B., Roatsch T., Denk T., Wolf U., Porco C.C. , 2010. The geology of Rhea: a first look at the ISS camera data from orbit 121 (Nov. 21 2009) in Cassini's extended mission. *Lunar Planet. Sci.* **41**, 1672

Zahnle K., Schenk P., Sobieszczyk S., Dones L., Levison H.F. 2001. Differential cratering of synchronously rotating satellites by ecliptic comets. *Icarus* **153**, 111-129

Zahnle K., Schenk P., Levison H., Dones L, 2003. Cratering rates in the outer Solar System. *Icarus* **163**, 263-289

# FIGURE 1

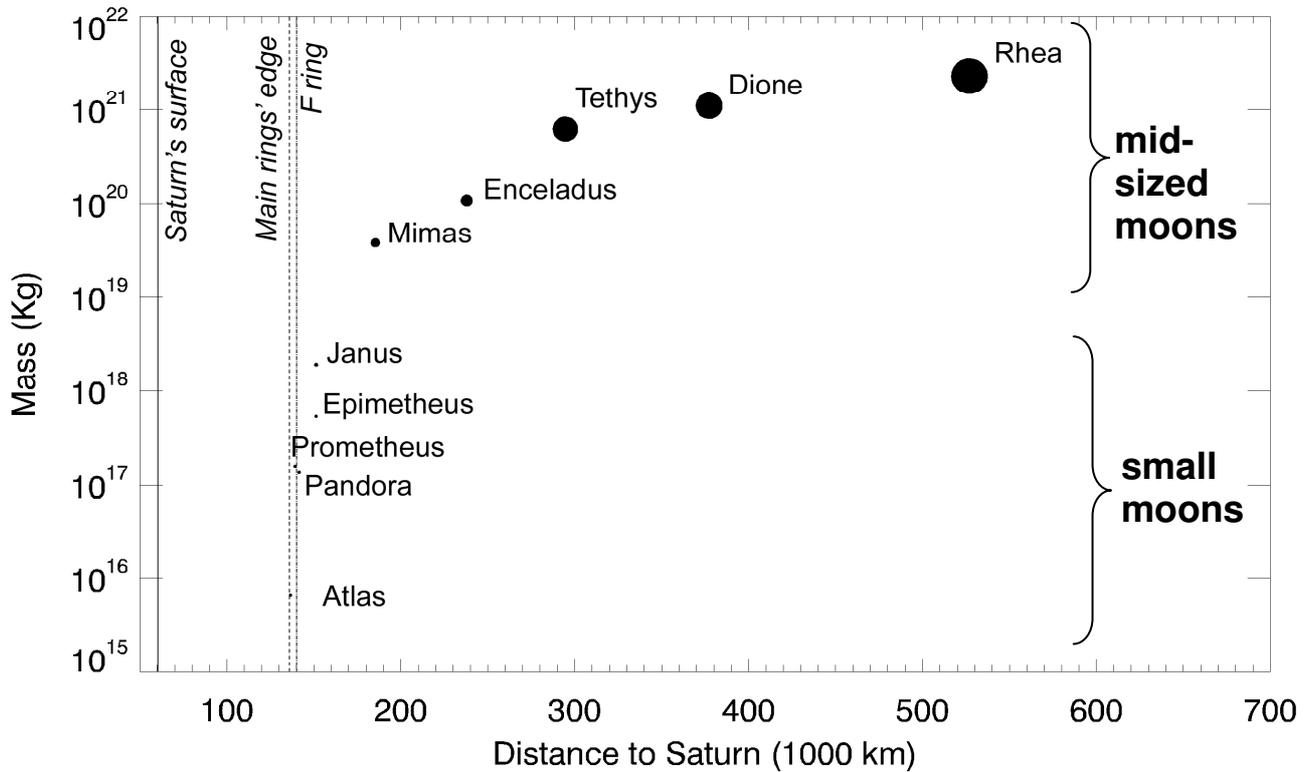

**Figure 1:** Mass of Saturn's icy inner moons versus distance. "Small moons" designate moons from Atlas to Janus, and "mid-sized moons" those from Mimas to Rhea. Circles' radii are proportional to the size of the moons. The vertical solid line shows Saturn's equatorial radius at 60,330 km, the dashed line shows the location of the outer edge of Saturn's A ring at 136,750km, and the vertical dashed-dotted lines stands for the F-ring.

**FIGURE 2**

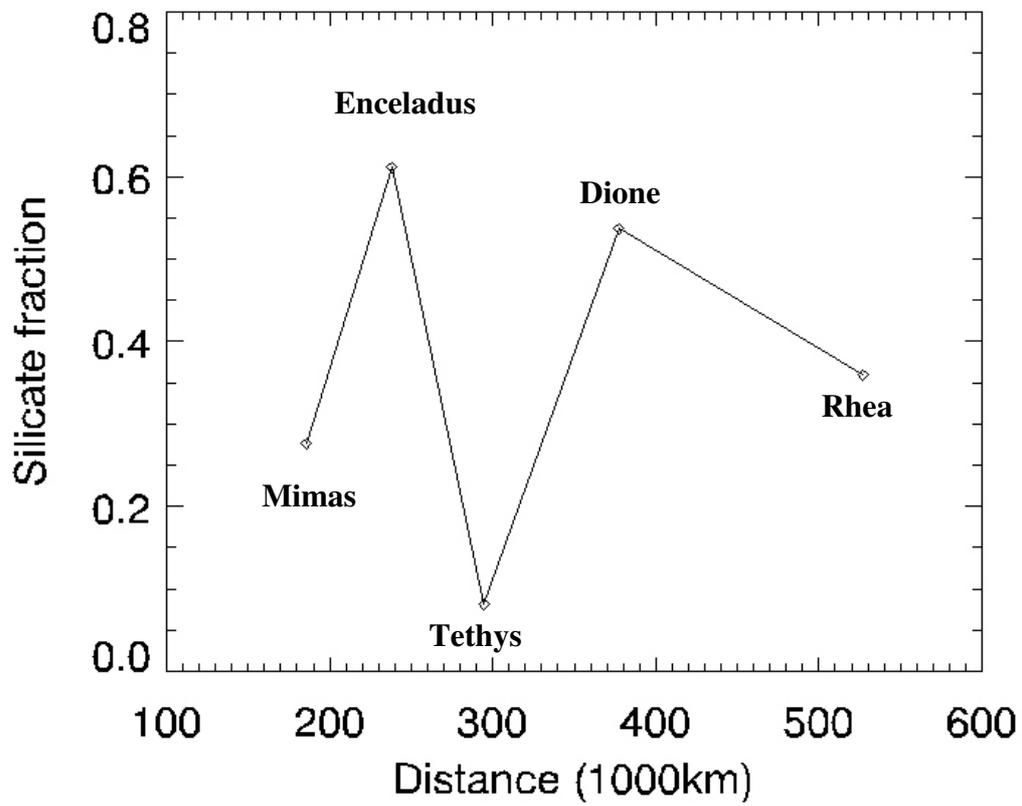

**Figure 2:** Silicate mass fraction of today's Saturn mid-sized moons. These values are computed assuming a silicate density of 3000 kg/m$^3$ and a solid ice density of 930 kg/m$^3$. For the least evolved of these objects (Mimas, Tethys, possibly Rhea), the silicate fraction may be underestimated due to the long-term preservation of remnant porosity.

# FIGURE 3

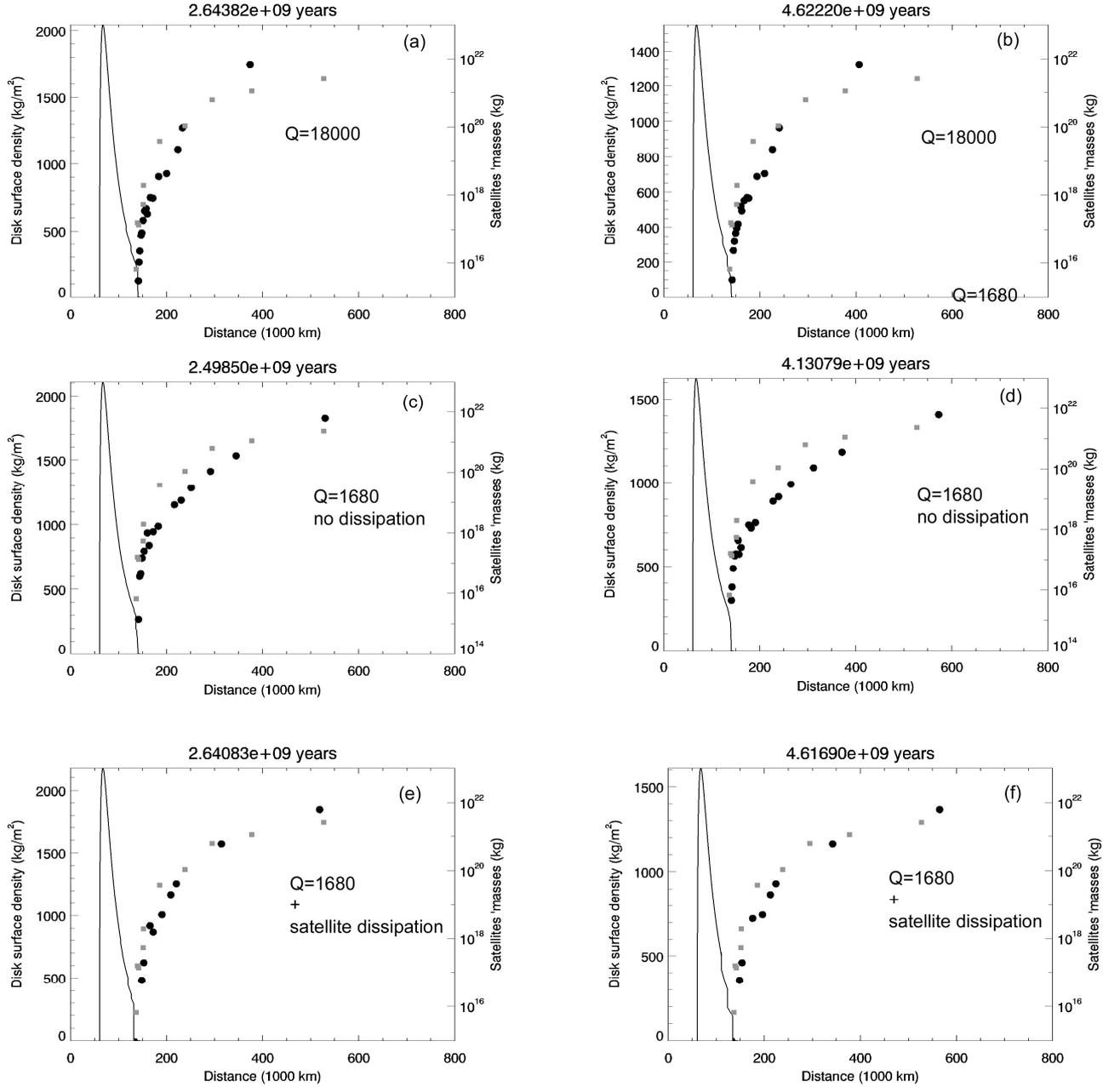

**Figure 3**: Evolution of the rings' surface density (solid line, left scale) and mass of proto-satellites (black solid dots, right scale) as a function of the distance to Saturn's center. Grey squares represent the current population of saturnian satellites. The initial disk mass is equal to 4 Rhea masses. In all simulation $k_{2p}=0.341$. The positions of the satellites formed from Saturn's rings are shown after 2.5Gyr and 4.5Gyr evolution. Two values of dissipation factor are tested for Saturn: $Q_p=18,000$ (a,b, Case A) and $Q_p=1680$ (c,d, Case B). The dissipation in the satellites is included assuming $k_2^s/Q_s=0.01$ for all satellites with $Q_p=1680$ (e and f, case C).

**FIGURE 4**

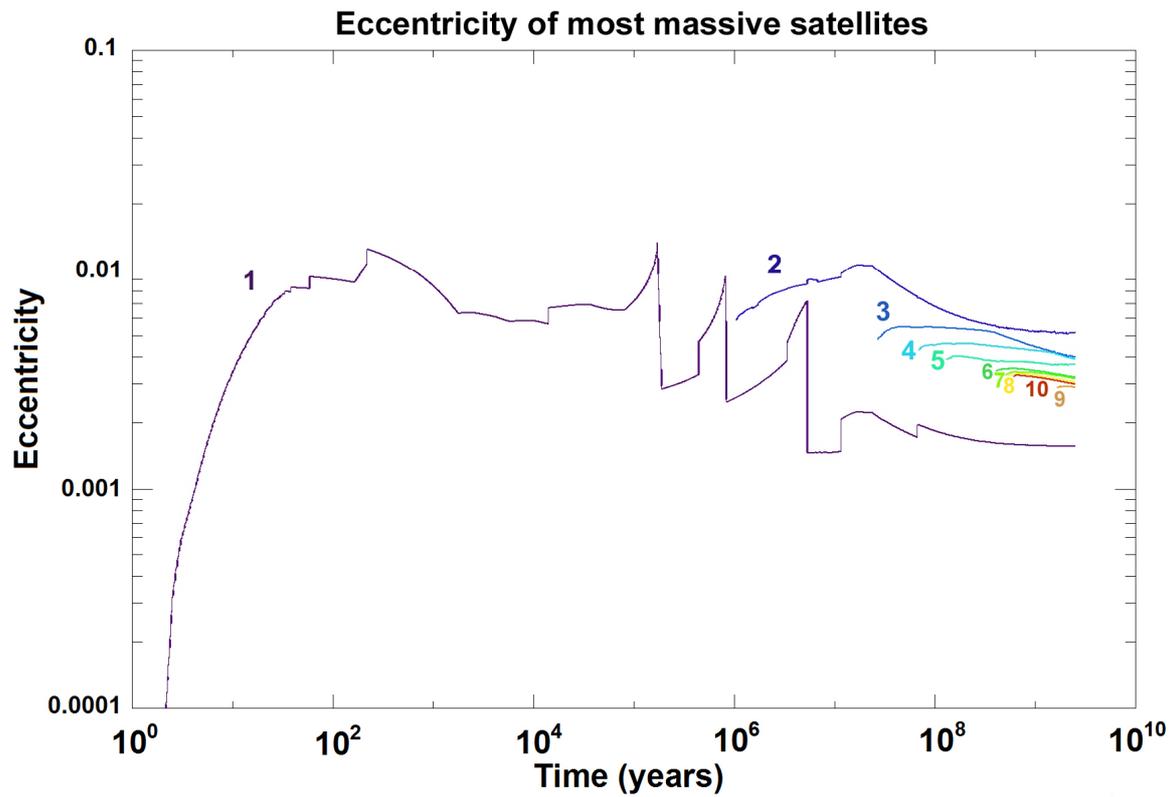

**Figure 4:** Satellites' eccentricity as a function of time in case C (see section 3). Colors and numbers indicate the rank of the satellite by decreasing order of mass at the end of the simulation ~ 4.7 Gyr ("1" is the most massive satellites, "2" is the second most massive one,…, "10" is the least massive satellite).

**FIGURE 5**

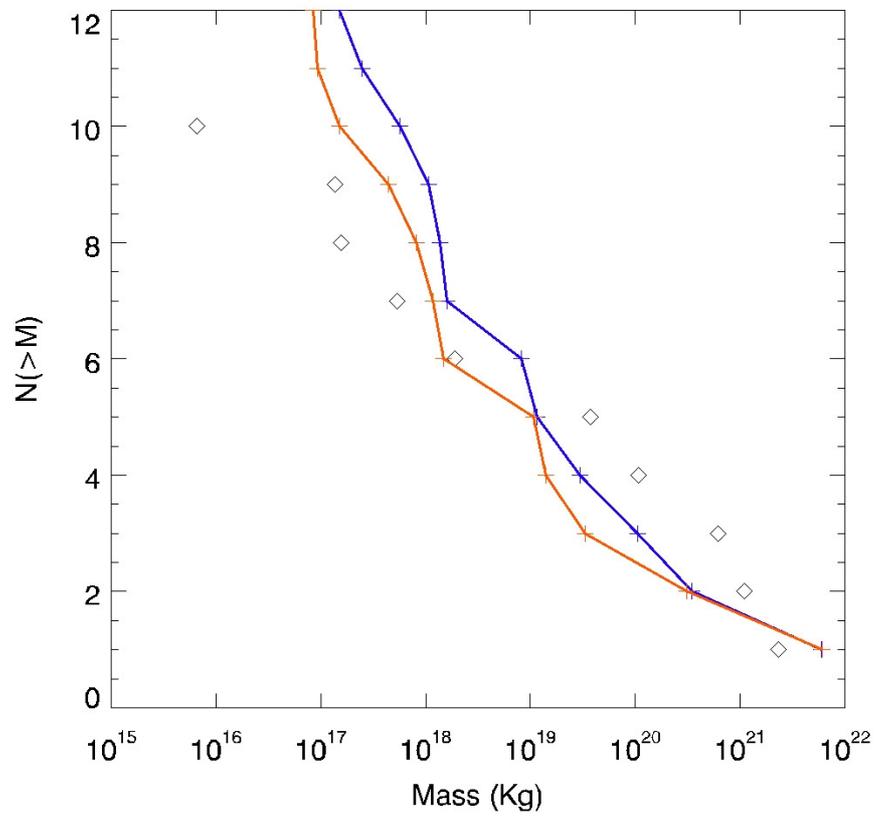

**Figure 5:** Cumulative mass distribution of moons formed in our simulations, assuming $Q_p=1680$. Red: simulations including satellite dissipation (Case C); blue: without satellite dissipation (Case B); black diamonds: current cumulative mass distribution of Saturn's satellites up to Rhea.

# FIGURE 6

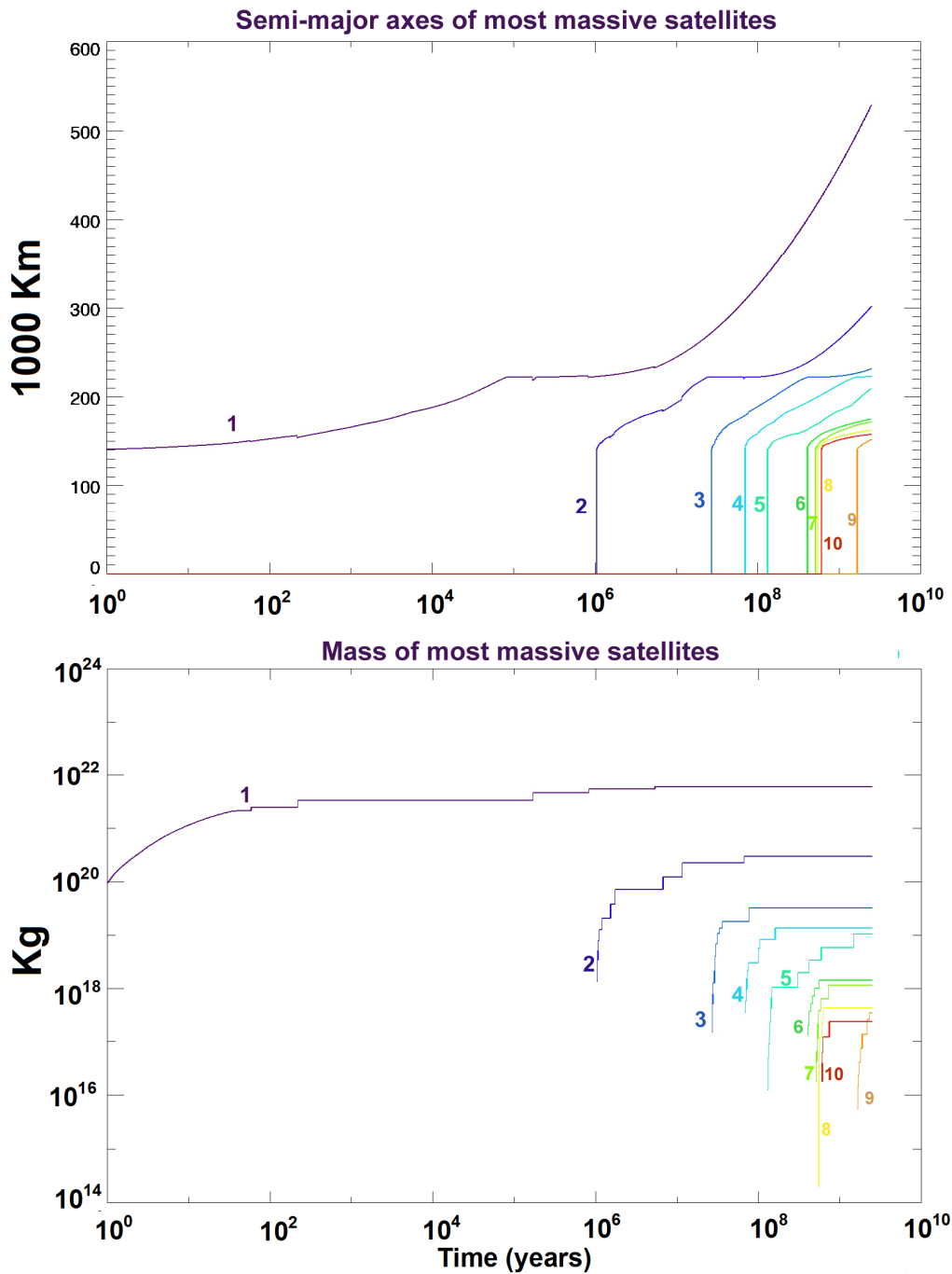

**Figure 6:** Orbital and mass evolution of satellites in a simulation with a $4M_R$ initial disk mass, assuming $Q_p=1680$ and satellite dissipation ( $k_2^s/Q_2^s =0.01$). Colors and numbers indicate the rank of the satellite by decreasing order of mass at the end of the simulation ~ 4.7 Gyr ("1" is the most massive satellites, "2" is the second most massive one,…, "10" is the least massive satellite).Top: Semi-major axes of the 10 most massive bodies as a function of time; bottom: mass of the 10 most massive satellites as a function of time. Conclusion: the less massive the object, the younger it is.

# FIGURE 7

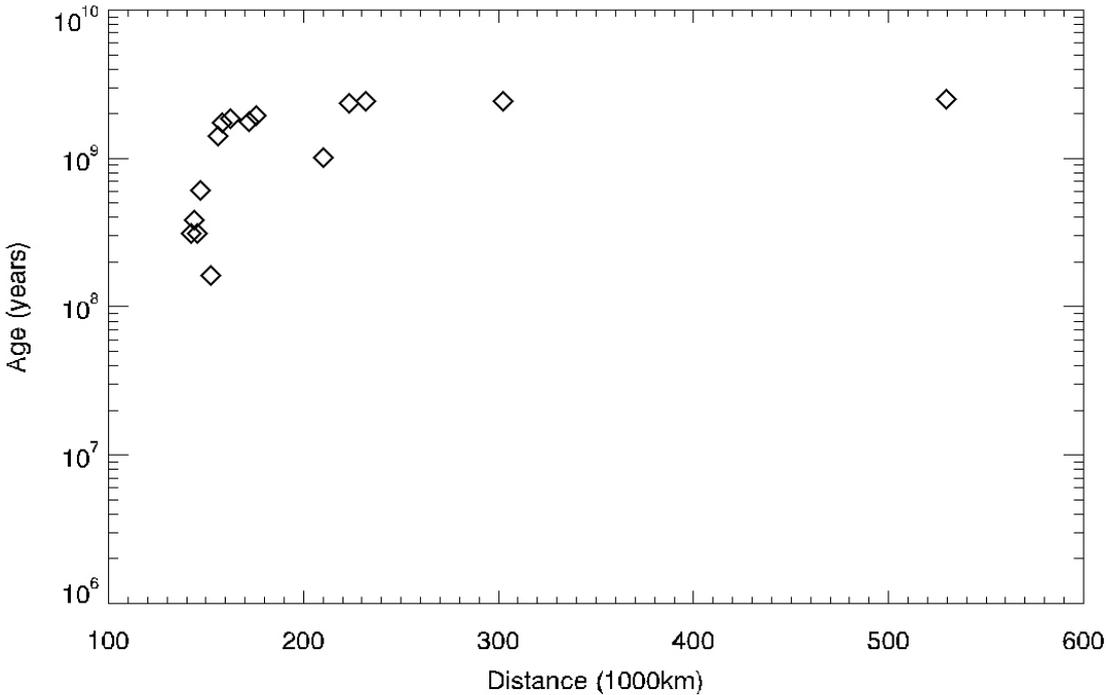

**Figure 7:** Age of the satellites assuming Saturn's rings formed 2.5 Gyr ago (Case C). The age of a satellite is referenced to the time at it acquired 80% of its final mass.

**Figure 8**

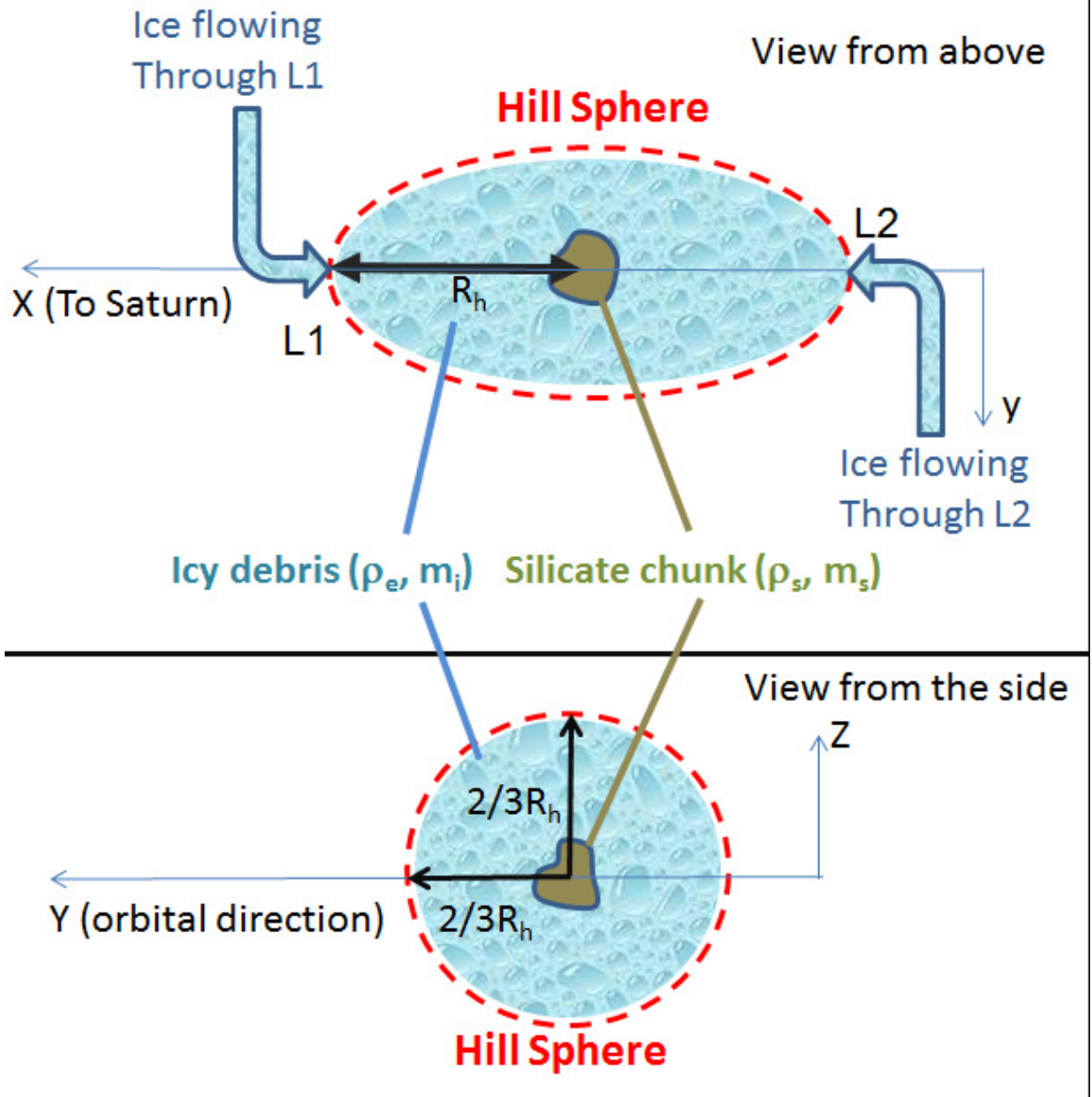

**Figure 8:** Sketch depicting the process of heterogeneous accretion of icy material onto a silicate chunk in Saturn's rings. The icy material flows to the silicate body through the L1 and L2 Lagrange points. The accretion stops when the body's Hill sphere is filled. The Hill Sphere is an ellipsoid with axes ratio 3:2:2 and with the longest axis equal to the Hill Radius ($R_h$). (Top) View from above; X is the radial axis, while Y is the orbital direction. (Bottom) front-view. The end-result of this process is the formation of a differentiated proto-moon.

# Figure 9

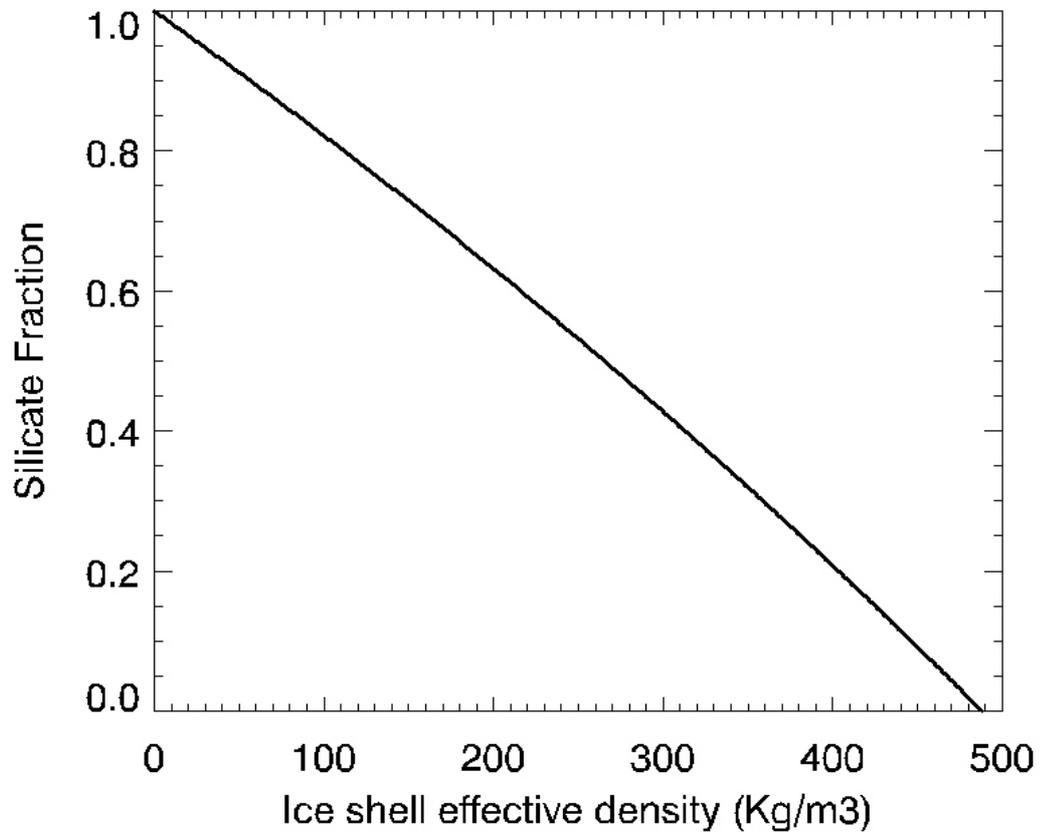

**Figure 9**: Silicate mass fraction ($m_s/(m_s+m_i)$) as a function the ice effective density ($\rho_e$) in the proto-moon's outer shell, in order to fill a body's Hill Sphere located at $a_s$=130,000 km.

**Figure 10**

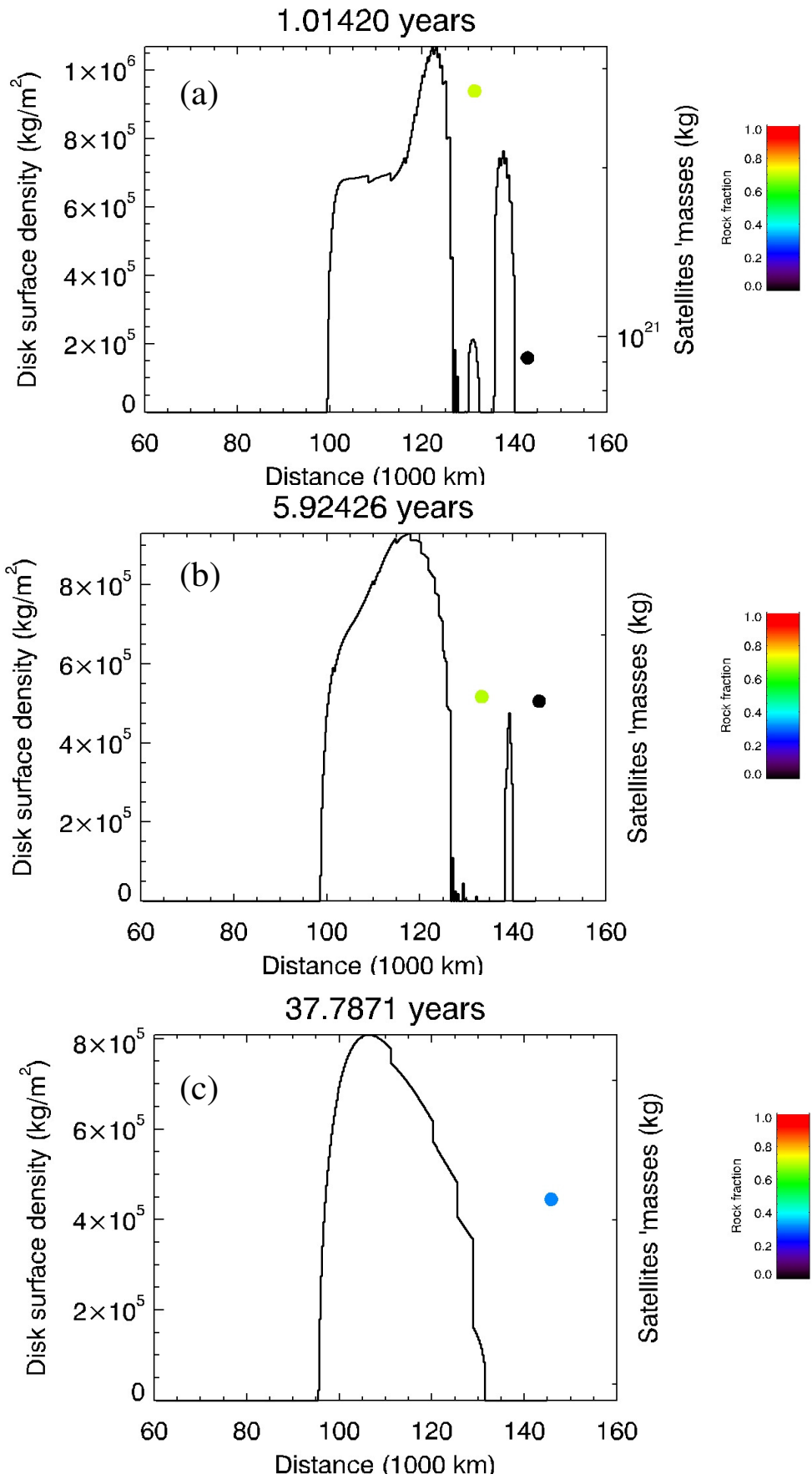

**Figure 10:** Evolution of a disk with 4 Rhea masses in which a large seed of pure silicate is embedded (with initial mass $2\times10^{21}$ kg), onto which icy ring material accretes (see Section 4.2.1 for details). The disk's surface density (solid line) is displayed on the left ordinate, the total mass of satellites formed from the ring material (solid circles) are displayed on the right ordinate. The silicate mass fractions of the newly formed moons are color-coded after the scale displayed on the right. The silicate proto-moon is initially implanted at 130,000 km. It rapidly accretes an icy shell (see Fig.11), and migrates outward as a consequence of tidal interactions (panels a-b). During the same timeframe, moons form at the ring's outer edge from pure water ice (see panel b). The two populations of moons are bound to meet and merge, as illustrated in panel c.

**FIGURE 11:**

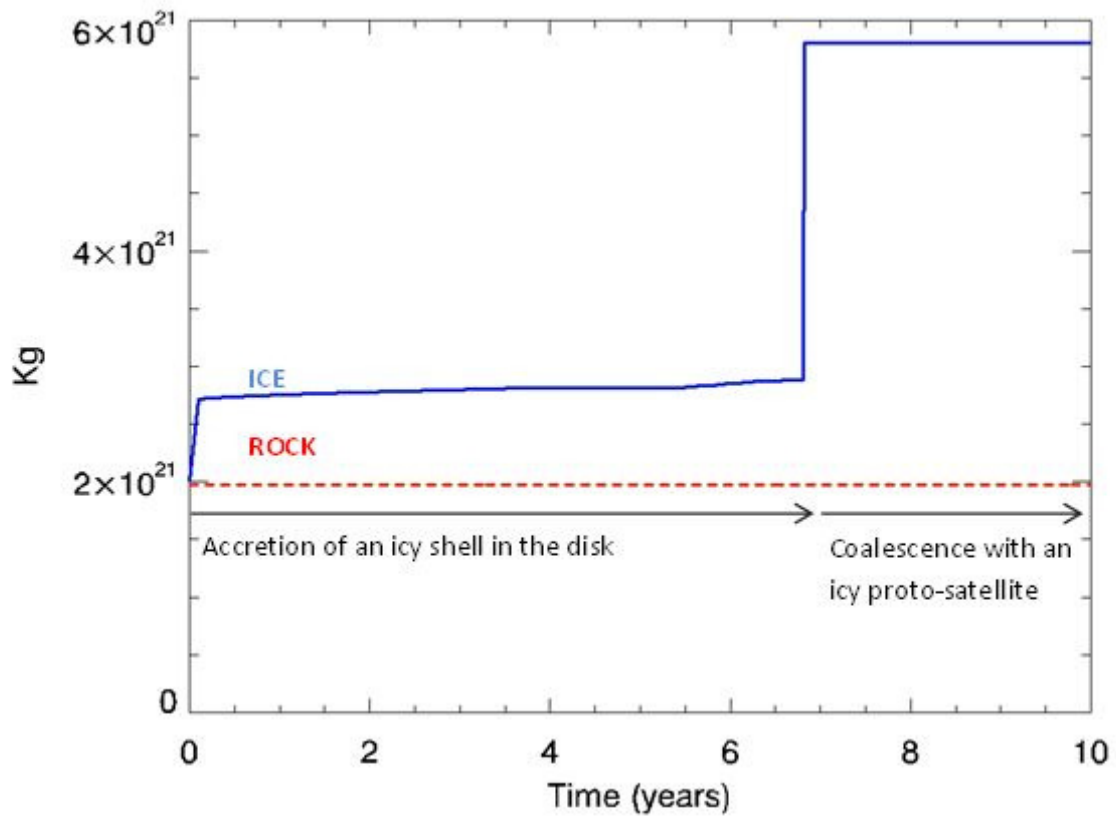

**Figure 11:** Mass evolution of a proto-moon initially made up of silicates, evolving in an icy disk (Fig. 10): from 0 to ~6.8 years a shell of icy material is accreted while the satellite remains inside the disk, which results in the opening of a gap (Figure 10). At 6.8 years the proto-moon leaves the disk through its outer edge and fuses with a nearby proto-satellite made of pure ice (visible in Fig.10b). The fusion results in a single object with ~30% silicate in mass. Blue line: total mass; red-dashed-line: mass of silicate core.

# FIGURE 12

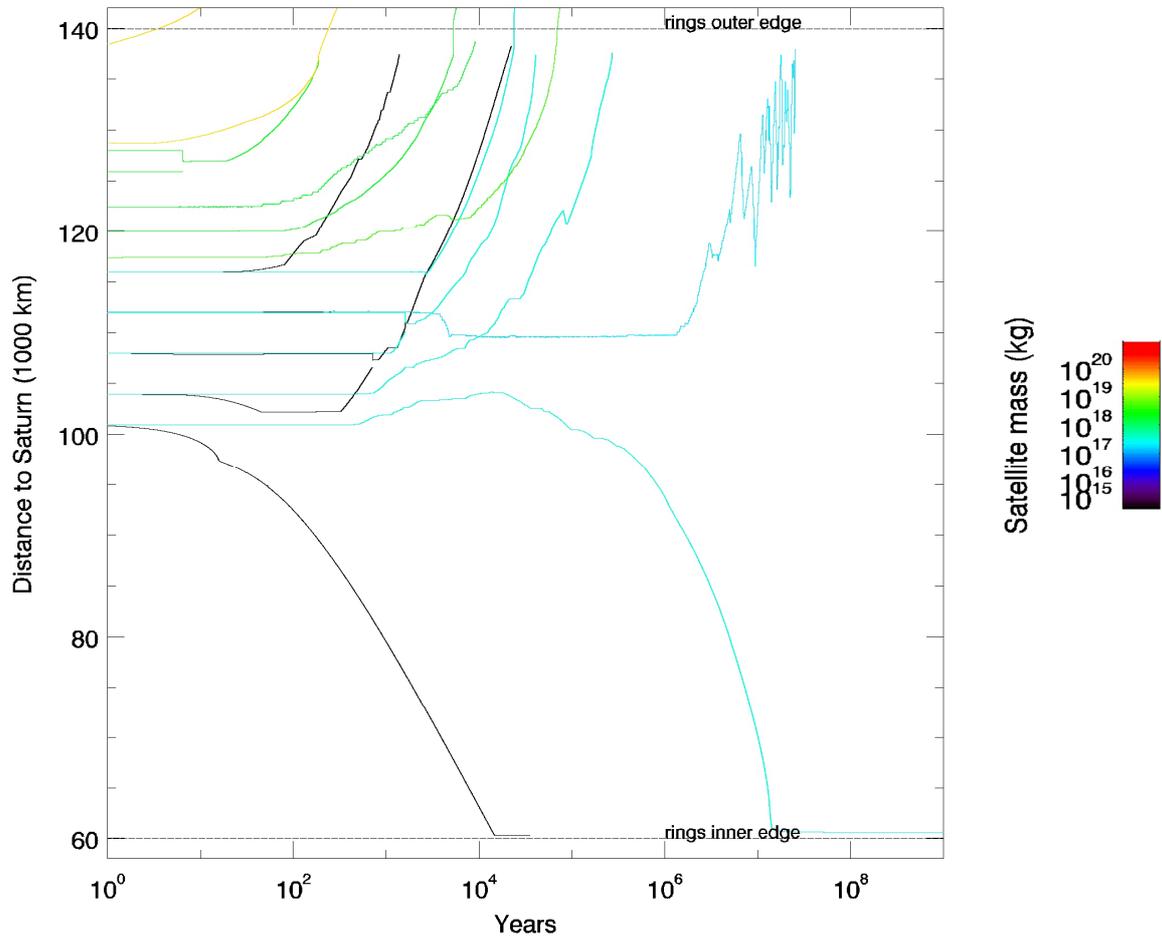

**Figure 12:** Evolution of semi-major axes for silicate satellites initially embedded in the ring, for a range of initial moon masses (right scale). The tidal interactions with the disk systematically leads to the ejection of the embedded satellites from the disk.

**APPENDIX: Tidal ages of the current Saturn's population of icy satellites**

Satellite evolution is dominated in a first stage by the torque from the rings. Then, after they reach the location of the outer 2:1 mean motion resonance with the edge of the rings ($a_{r21} = 2^{2/3} * R_L = 222\ 236$ km), this torque vanishes, and the tides from Saturn dominate. When they reach $a_{r21}$, the satellites acquire their final mass (see Section 3.4 and Fig. 6). From Eq. (1), taking into account only the first term of the right-hand side (that is, the term relative to the tides from Saturn), and assuming a constant mass for the satellite, we derive the following relationship between the mass of a satellite and its semi-major axis, expressed as a function of the time $t$ since it reached $a_{r21}$:

$$m_s = (a_s^{13/2} - a_{r21}^{13/2}) / (C\, t) \qquad \textbf{Eq. (A1)}$$

where $C = (39/2)\, k_{2p}\, G^{1/2}\, R_p^5 / Q_p\, M^{1/2}$. This relation defines isochrones: all the satellites that fall on the curve $m_s(a_s)$ defined by Eq.A1 for a given $t$ and value of $a_{r21}$ were, in the past, at distance $a_{r21}$ from Saturn at time $t$ ago. Put differently, using graphics, if we impose $t$, Eq.(A1) defines a curve in the mass – semi-major axis diagram, that is an isochrone : all satellites with different masses but with a same date and location of birth will evolve in the mass – semi-major axis diagram on a same curve as defined by Eq.A1. Several isochrones are plotted on Figure A1 assuming $Q_p=1680$ and $k_{2p}=0.341$. Conversely, one can estimate the "tidal ages" of the satellites, by inverting this relation and finding $t$ for given $m_s$ and $a_s$ (see Figure A2).. We see that the tidal ages are an increasing function of the semi-major axis, for the saturnian satellites located beyond $a_{r21}$. This means that it is possible that they formed one after another and then migrated outward, like in the scenario we propose here. In this case, the outermost satellites are older than the innermost ones. This chronology may be tested against more detailed geological evolution models. The ages yielded by our modeling (Fig. A2) indicate that all satellites up to Rhea could be formed by this process over the age of the solar system, while Titan could not (assuming a constant $Q_p=1680$ in time and space).

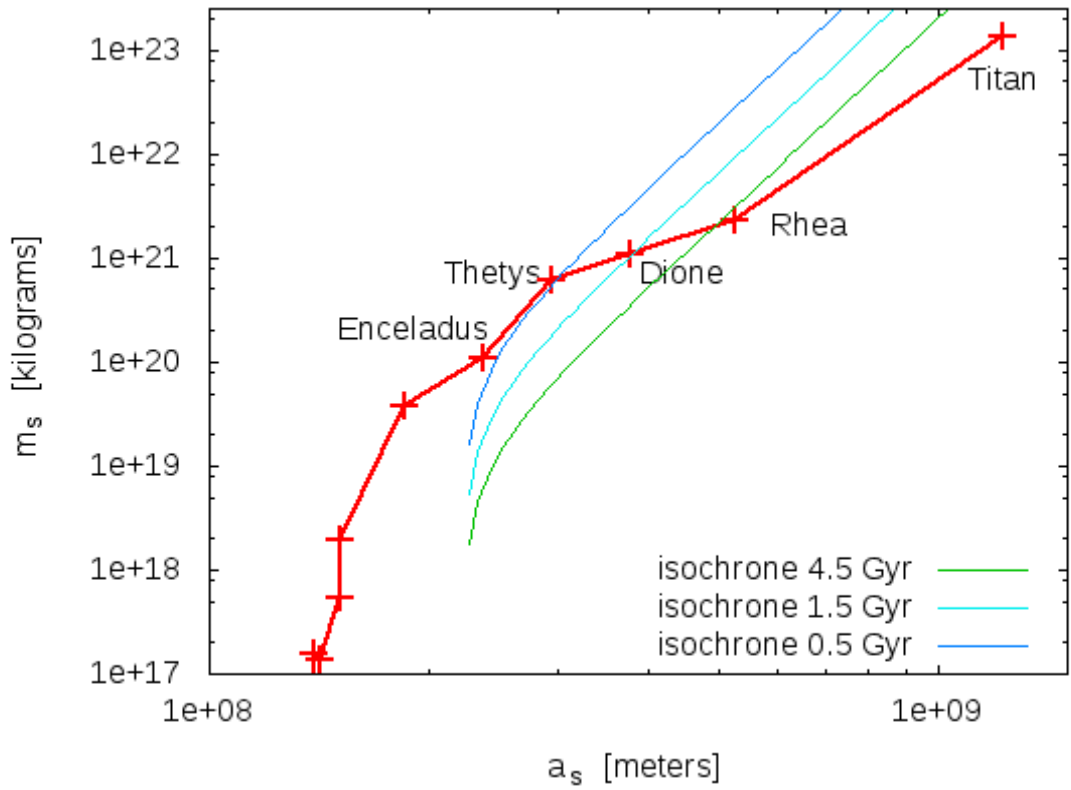

**Figure A1:** Mass versus semi-major axis diagram of the satellites of Saturn today (red dots and solid line). The dashed lines show isochrones defined by Eq.(A1) for various values of *t*. We see for instance that the time needed for Dione to migrate to its present semi-major axis (through Saturn's tides) since it acquired its final mass (at $a_{r21}$=220000km) is 1.5 Gyr.

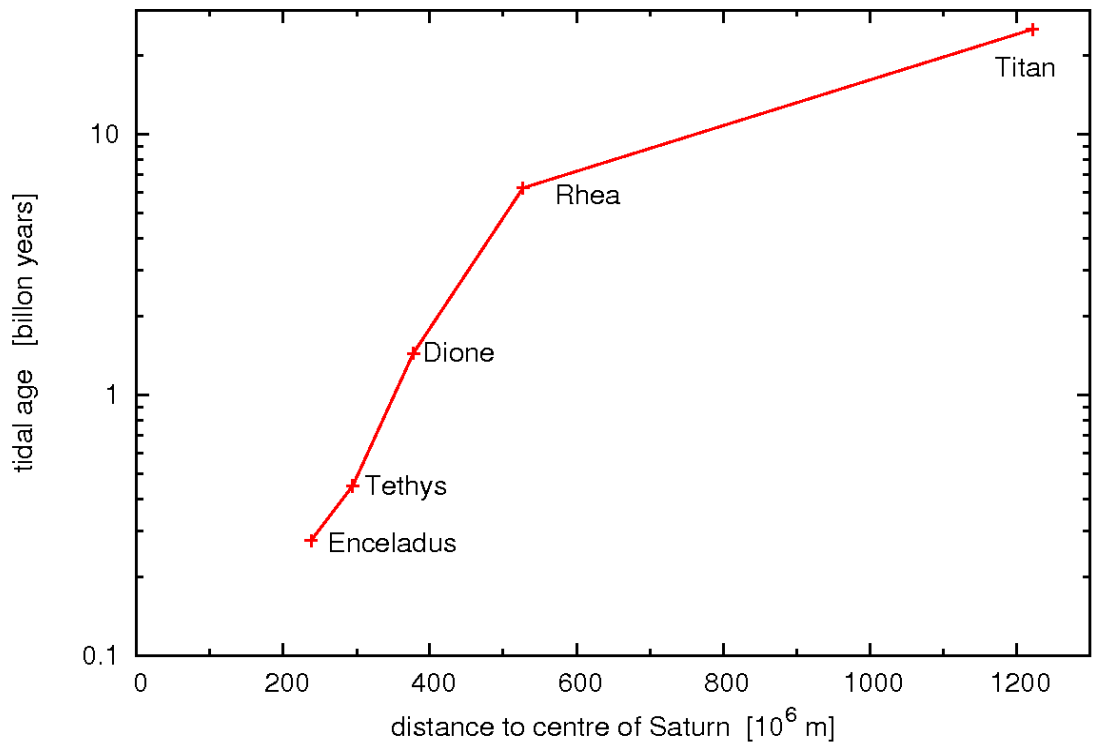

**Figure A2:** Tidal ages of the satellites beyond $a_{r21}$.